\documentclass[prd,reprint,longbibliography,nofootinbib,superscriptaddress]{revtex4-2}

\usepackage[USenglish]{babel}
\usepackage[T1]{fontenc}
\usepackage{amsmath}
\usepackage{mathtools}
\usepackage{amssymb}
\usepackage{dsfont}
\usepackage{graphicx}
\usepackage{physics}
\usepackage[acronym]{glossaries}
\usepackage{siunitx}
\usepackage[svgnames]{xcolor}
\usepackage[hidelinks]{hyperref}
\hypersetup{
	colorlinks = true, 
	urlcolor = DarkBlue, 
	linkcolor = DarkBlue, 
	citecolor = DarkBlue  
}
\usepackage{cleveref}
\creflabelformat{equation}{#2\textup{#1}#3}

\begin{document}
	
\title{Clock synchronization and light-travel-time estimation for space-based gravitational-wave detectors}
	
\author{Jan Niklas Reinhardt}
\email{janniklas.reinhardt@aei.mpg.de}
\affiliation{Max-Planck-Institut für Gravitationsphysik (Albert-Einstein-Institut),\\ Callinstraße 38, 30167 Hannover, Germany}
\affiliation{Leibniz Universität Hannover, Welfengarten 1, 30167 Hannover, Germany}

\author{Olaf Hartwig}
\affiliation{Max-Planck-Institut für Gravitationsphysik (Albert-Einstein-Institut),\\ Callinstraße 38, 30167 Hannover, Germany}
\affiliation{Leibniz Universität Hannover, Welfengarten 1, 30167 Hannover, Germany}

\author{Gerhard Heinzel}
\affiliation{Max-Planck-Institut für Gravitationsphysik (Albert-Einstein-Institut),\\ Callinstraße 38, 30167 Hannover, Germany}
\affiliation{Leibniz Universität Hannover, Welfengarten 1, 30167 Hannover, Germany}


\begin{abstract}
    Space-based gravitational-wave detectors, such as LISA, record interferometric measurements on widely separated satellites. Their clocks are not synced actively. Instead, clock synchronization is performed in on-ground data processing. It relies on measurements of the so-called pseudoranges, which entangle the interspacecraft light travel times with the clock desynchronizations between emitting and receiving spacecraft. For interspacecraft clock synchronization, we need to isolate the differential clock desynchronizations, i.e., disentangle the pseudoranges. This further yields estimates for the interspacecraft light travel times, which are required as delays for the laser frequency noise suppression via time-delay interferometry. Previous studies on pseudorange disentanglement apply various simplifications in the pseudorange modeling and the data simulation. In contrast, this article derives an accurate pseudorange model in the barycentric celestial reference system, complemented by realistic state-of-the-art LISA data simulations. Concerning pseudorange disentanglement, this leads to an a priori under-determined system. We demonstrate how on-ground orbit determinations, as well as onboard transmission and on-ground reception time tags of the telemetry data, can be used to resolve this degeneracy. We introduce an algorithm for pseudorange disentanglement based on a nonstandard Kalman filter specially designed for clock synchronization in systems where pseudorange measurements are conducted in different time frames. This algorithm achieves interspacecraft clock synchronization and light travel time estimation with submeter accuracy, thus fulfilling the requirements of time-delay interferometry. 
\end{abstract}

\maketitle

\section{Introduction}\label{sec:Introduction}
On September 14, 2015, the Laser Interferometer Gravitational-Wave Observatory (LIGO) accomplished the first direct detection of gravitational waves \cite{FirstGwDetection}. From that day on, many gravitational-wave signals have been detected by ground-based gravitational-wave detectors \cite{GravitationalWaveTransientCatalogFirstAndSecondRun,GravitationalWaveTransientCatalogThirdRun1,GravitationalWaveTransientCatalogThirdRun2}. Yet, due to the gravity gradient noise and the limited extent of interferometer arms on ground, these detectors are blind for gravitational waves in the sub-hertz frequency band, where various interesting sources are lurking \cite{Amaro:AstrophysicsWithLISA}. To circumvent these difficulties, several space-based gravitational-wave detectors are being planned \cite{Taiji,Tianqin,Decigo}. One of these is the Laser Interferometer Space Antenna (LISA) aiming for the frequency band between \SI{0.1}{\milli \hertz} and \SI{1}{\hertz} \cite{LISA}. LISA was recently adopted by the European Space Agency (ESA) and is expected to be launched around the year 2035.\par
LISA consists of three spacecraft (SC) on heliocentric geodesics trailing or leading Earth by about 20 degrees. They constitute an approximate equilateral triangle with an arm length of about 2.5 million \si{\kilo\m}, which completes one cartwheel rotation per orbit. To keep this triangle as rigid and equilateral as possible, the constellation plane is tilted about 60 degrees out of the ecliptic \cite{Nayak:MinimumFlexingArms}. Each SC houses two lasers with a nominal wavelength of \SI{1064}{\nano\m}, which are transmitted to the other two SC of the constellation. Heterodyne interferometry between received and local lasers yields six interspacecraft beat notes at \SIrange{5}{25}{\mega\hertz}, which are detected by quadrant photoreceivers. Gravitational waves cause picometer variations in the interspacecraft distances, showing up as phase fluctuations in these beat notes. Phasemeters are applied to extract the beat note phases with microcycle precision \cite{GerberdingPhasemeter}.\par
However, various instrumental noise sources swamp the phasemeter output, thus burying the gravitational-wave signals. The dominant noise source is the laser frequency noise exceeding the target sensitivity by more than 8 orders of magnitude. A further challenge is the timing: Each SC houses an ultrastable oscillator (USO), from which an \SI{80}{\mega \hertz} clock signal is coherently derived. The associated counter is considered the onboard timing reference and referred to as spacecraft elapsed time (SCET). All interferometric measurements are timestamped according to the SCET of the satellite where they are taken. The SCETs deviate from each other and from the barycentric coordinate time (TCB) due to relativistic effects and instrumental drifts and jitters. Hence, the LISA measurements are a priori not synchronized.\par
Laser frequency noise can be mitigated in on-ground data processing via time-delay interferometry (TDI) \cite{ArmstrongTDI,TintoTDIforLISA}. TDI linearly combines appropriately delayed versions of the various beat notes to virtually form equal-optical-path-length interferometers, in which laser frequency noise naturally cancels. In the baseline topology, TDI is performed after synchronizing the interferometric measurements to some universal time frame, e.g., TCB. The required delays are the interspacecraft light travel times. In a recently proposed alternative topology, TDI is performed before clock synchronization using the pseudoranges as delays \cite{Hartwig:TDIwoSync}. Here, the laser-noise-free TDI combinations are computed in the respective SCETs and must be synchronized to TCB a posteriori.\par
The TDI baseline topology requires the interspacecraft light travel times, which are used as delays, and the differential SCET desynchronizations for the preceding measurement synchronization (see \cref{fig:PseudorangeDisentanglementInTDIBaselineTopology}). However, these quantities are not directly measured in LISA. Instead, the onboard pseudorandom noise (PRN) ranging scheme measures pseudoranges \cite{Heinzel-Ranging}: The pseudorange is commonly defined as the difference between the SCET of the receiving SC at reception and the SCET of the emitting SC at emission. Thus, it represents a combination of the geometrical range (light travel time) and the differential SCET desynchronization. To synchronize the SCETs with each other and to obtain the light travel times for the TDI baseline topology, we need to disentangle the pseudoranges and isolate these constituents.\par
Previous studies successfully applied a Kalman filter (KF) to disentangle the pseudoranges \cite{Wang:KF-I, Wang:KF-II, Wang:Thesis}. However, those investigations considered a simplified LISA model: (1) The light travel times of counterpropagating laser links were assumed to be equal. Thus, the six pseudoranges could be expressed via three arm lengths and two relative SCET desynchronizations, which led to a well-defined system. Realistic modeling in the barycentric celestial reference system (BCRS) requires different light travel times for counterpropagating laser links due to the heliocentric motion. Even in a LISA-centered reference frame, the light travel times are not symmetric due to the Sagnac effect associated with the cartwheel rotation \cite{Shaddock:SagnacLISA}. (2) The SCET deviations of emitting and receiving SC were both considered at the same time, i.e., the light travel times were partially neglected in the modeling. (3) In \cite{Wang:KF-I,Wang:KF-II}, the desynchronization was neglected in the measurement timestamps: While the relative SCET desynchronizations were incorporated as pseudorange constituents, all measurements were assumed to be recorded according to some universal time frame. A prototype implementation of distorted measurements was briefly considered in \cite{Wang:Thesis}, but that investigation needs to be elaborated.
(4) The LISA data was simulated using Keplerian orbits and a non-relativistic framework.\par
Apart from these simplifications, the previous studies mainly focused on KF state vectors, including absolute quantities. Such state vectors cannot be determined by the LISA interferometric measurements, which only measure relative quantities, and their dimension highly exceeds the number of available measurements. Moreover, LISA models based on absolute positions involve substantial nonlinearities, resulting in numerical instabilities. Treating absolute positions in the BCRS along with clock variables further leads to state vectors with a dynamic range of around 18 orders of magnitude.\par
It was recently demonstrated how the four LISA ranging observables could be combined to obtain accurate and precise pseudorange estimates \cite{Reinhardt:RangingSensorFusion}. These estimates build the starting point of the investigations conducted in this article. In \cref{sec:ClockSynchronizationViaPseudorangeDisentanglement}, we derive a pseudorange expression in the BCRS without the simplifications (1) and (2). We point out that dropping these assumptions leads to an under-determined system. To resolve the degeneracy, we suggest including additional measurements in the pseudorange disentanglement process. These measurements are the orbit determinations and the mission operation center (MOC) time correlations introduced in \cref{sec:OrbitDetermination,sec:MocTimeCorrelation}. In \cref{sec:PseudorangeDisentanglement}, we present an algorithm for pseudorange disentanglement based on the iterative application of a modified KF. This nonstandard KF is particularly designed for clock synchronization in systems where ranging measurements are given in different time frames. We present the performance of the algorithm in \cref{sec:Results} using data simulated without the simplifications (3) and (4) and conclude in \cref{sec:Conclusion}.

\section{Clock Synchronization via Pseudorange Disentanglement}\label{sec:ClockSynchronizationViaPseudorangeDisentanglement}
Clock synchronization requires a convention of simultaneity. We consider the coordinate synchronization convention: Two clocks $\tau_1$ and $\tau_2$ are synchronized in some coordinate system $(t,\:\vec{x})$ if and only if $\tau_1(t) = \tau_2(t)$ for all relevant coordinate times $t$. In practice, this means that we need to determine the desynchronization
\begin{align}
    \delta \tau_{12}(t) = \tau_1(t) - \tau_2(t),
\end{align}
between both clocks in the chosen coordinate system.\par 
Here, we choose the barycentric celestial reference system (BCRS), which is the reference frame of LISA data analysis: Positions in the BCRS are expressed with respect to the barycenter of the solar system. The associated barycentric coordinate time (TCB) is the proper time corresponding to a clock at rest in a reference frame comoving with the barycenter of the solar system but outside of its gravity field.\par
Following \cite{Bayle:LISA-Simulation}, we denote the TCB by $t$, the SC proper times by $\tau_i$ (we just need them in \cref{sec:Data-simulation} to describe the data simulation), and the SCETs by $\hat{\tau}_{i}$. We add a superscript to indicate that a quantity is expressed as a function of a certain time frame, e.g., $\hat{\tau}_{2}^{\hat{\tau}_{1}}$ denotes the SCET of SC 2 as a function of the SCET of SC 1. The SCETs can be expressed in TCB via
\begin{align}\label{eq:TimeFrameDeviation}
    \hat{\tau}_{i}^{t}(x) = x + \delta \hat{\tau}_{i}^{t}(x),
\end{align}
where $\delta \hat{\tau}_{i}^{t}$ denotes the SCET desynchronization from TCB and $x$ is a generic function argument. $\delta \hat{\tau}_{i}^{t}$ comprises the deviation of the SCET from the corresponding SC proper time due to instrumental drifts and jitters, and the deviation of that proper time from TCB. Note that for any time frame $a$, we have
\begin{align}
    a^a(x) = x,\label{eq:TimeFrameInSameTimeFrame}\\
    \delta a^a(x) = 0.\label{eq:TimeFrameDeviationFromSameTimeFrame}
\end{align}
We use the double index notation for interspacecraft signal propagation, where the first index refers to the receiving SC and the second index to the emitting one, e.g., $R_{12}$ denotes the pseudorange measured on SC 1 for a signal incoming from SC 2.

\subsection{The pseudorange in the BCRS}\label{sec:PseudorangeInBCRS}
Having established the coordinate synchronization convention in the BCRS as our convention of simultaneity, we now need to express the pseudoranges in the BCRS. The pseudorange is defined as the difference between the SCET of the receiving SC at the event of reception and the SCET of the emitting SC at the event of emission \cite{HartwigThesis}. We express both SCETs as functions of TCB and evaluate them at the events of reception and emission, respectively. Taking the difference yields a first expression for the pseudorange in the BCRS:
\begin{align}
    R_{ij}^t(x_\text{rec})= \hat{\tau}_{i}^{t}(x_{\text{rec}}) - \hat{\tau}_{j}^{t}(x_{\text{emit}}).
\end{align}
Writing the emission TCB as the difference between the reception TCB and the interspacecraft light travel time (LTT), denoted by $d_{ij}^{t}$, gives
\begin{align}\label{eq:PseudorangeInTCB-Temp-I}
    R_{ij}^t(x_\text{rec})=\hat{\tau}_{i}^{t}(x_{\text{rec}}) - \hat{\tau}_{j}^{t}\left(x_{\text{rec}} - d_{ij}^{t}(x_{\text{rec}})\right).
\end{align}
We drop the subscript \textit{rec}. Below, $x$ refers to the TCB of reception. Applying \cref{eq:TimeFrameDeviation} to \cref{eq:PseudorangeInTCB-Temp-I} leads to
\begin{align}
    R_{ij}^t(x)&= \delta\hat{\tau}_{i}^{t}(x) + d_{ij}^{t}(x) - \delta \hat{\tau}_{j}^{t}\left(x - d_{ij}^{t}(x)\right).
\end{align}
We now expand the deviation of the SCET of the emitting SC from TCB, $\delta \hat{\tau}_{j}^{t}$, around zero LTT:
\begin{align}\label{eq:PseudorangeInTCB-Temp-II}
    R_{ij}^t(x)&= \delta\hat{\tau}_{i}^{t}(x) + d_{ij}^{t}(x)\nonumber\\ 
    - \Big( \delta &\hat{\tau}_{j}^{t}(x) - \delta \dot{\hat{\tau}}_{j}^{t}(x) \cdot  d_{ij}^{t}(x) + \frac{1}{2}\delta\ddot{\hat{\tau}}_{j}^{t} \cdot \left(d_{ij}^{t}(x)\right)^2 + ...\Big).
\end{align}
The relative SCET desynchronization and the LTT amount to \SIrange{1}{10}{\s}. According to relativistic modeling of the LISA orbits and analyses of space-qualified USOs, the linear and quadratic SCET drifts from TCB are expected to be in the order of $\delta\dot{\hat{\tau}}_{j}^{t} \propto 10^{-7}$ and $\delta\ddot{\hat{\tau}}_{j}^{t} \propto 10^{-14}\si{\s\tothe{-1}}$ \cite{Bayle:LisaOrbits,Asmar:USO}. Hence, the first-order term in the expansion contributes to the pseudorange in the order of \SI{1}{\micro\s} corresponding to a few hundred meters and needs to be considered. The second-order term contributes in the order of \SI{0.1}{\pico\s} (a few tenths of a millimeter) and can be neglected. The pseudorange can now be expressed in the BCRS via the LTT $d_{ij}^{t}$, the differential SCET desynchronization $\delta\hat{\tau}_{ij}^{t}$, and the SCET drift of the emitting SC with respect to TCB $ \delta \dot{\hat{\tau}}_{j}^{t}$:
\begin{align}
    R_{ij}^t(x) =&\: \delta\hat{\tau}_{ij}^{t}(x) + \left(1 + \delta \dot{\hat{\tau}}_{j}^{t}(x)\right) \cdot  d_{ij}^{t}(x),\label{eq:PseudorangeTCB}\\
    \delta\hat{\tau}_{ij}^{t}(x) \coloneqq&\: \delta\hat{\tau}_{i}^{t}(x) - \delta \hat{\tau}_{j}^{t}(x).\label{eq:DefinitionDifferentialScetOffset}
\end{align}

\subsection{Pseudorange disentanglement}
Interspacecraft clock synchronization in LISA requires information about the two differential SCET desynchronizations $\delta \hat{\tau}_{12}$ and $\delta \hat{\tau}_{13}$.\footnote{The desynchronization between SCET 2 and SCET 3 can be expressed via $\delta \hat{\tau}_{23} = \delta \hat{\tau}_{13} - \delta \hat{\tau}_{12}$ (see \cref{eq:DefinitionDifferentialScetOffset}).} Above, we demonstrate that these are constituents of the pseudoranges when expressed in a universal time frame like TCB. If we successfully isolate them in \cref{eq:PseudorangeTCB}, we can synchronize the three SCETs among each other.\par
Pseudorange disentanglement is challenging, though: The six pseudoranges are composed of six LTTs and two relative SCET desynchronizations. Moreover, the absolute SCET drifts with respect to TCB appear in \cref{eq:PseudorangeTCB} due to the expansion we performed in \cref{eq:PseudorangeInTCB-Temp-II}.\footnote{Previous studies \cite{Wang:KF-I, Wang:KF-II, Wang:Thesis} applied a simplified pseudorange model considering both SCET deviations at the same time. Hence, the expansion of \cref{eq:PseudorangeInTCB-Temp-II} was not required, and the SCET drifts did not appear in the equations.} It is possible to express the three absolute SCET drifts from TCB in terms of two differential SCET drifts (time derivatives of the two differential SCET desynchronizations) and one absolute SCET drift from TCB $\delta\dot{\hat{\tau}}_{1}^{t}$: \footnote{Similar equations can be derived considering $\delta\dot{\hat{\tau}}_{2}^{t}$ or $\delta\dot{\hat{\tau}}_{3}^{t}$ as absolute SCET drifts. It is convenient to have all three models due to the alternating LISA communication schedule (see \cref{sec:MocTimeCorrelation}).}
\begin{align}\label{eq:AllPseudoranges}
    \begin{bmatrix}
        R^t_{12}\\
        R^t_{23}\\
        R^t_{31}\\
        R^t_{13}\\
        R^t_{32}\\
        R^t_{21}
    \end{bmatrix}& = 
    \begin{bmatrix}
        \delta\hat{\tau}_{12}^{t} + (1 + \delta \dot{\hat{\tau}}_{1}^{t} - \delta \dot{\hat{\tau}}_{12}^{t}) \cdot d_{12}^t\\
        \delta\hat{\tau}_{13}^{t} - \delta\hat{\tau}_{12}^{t} + (1 + \delta \dot{\hat{\tau}}_{1}^{t} - \delta \dot{\hat{\tau}}_{13}^{t}) \cdot d_{23}^t\\
        - \delta\hat{\tau}_{13}^{t} + (1 + \delta \dot{\hat{\tau}}_{1}^{t}) \cdot d_{31}^{t}\\
        \delta\hat{\tau}_{13}^{t} + (1 + \delta \dot{\hat{\tau}}_{1}^{t} - \delta \dot{\hat{\tau}}_{13}^{t}) \cdot d_{13}^t\\
        \delta\hat{\tau}_{12}^{t} - \delta\hat{\tau}_{13}^{t} + (1 + \delta \dot{\hat{\tau}}_{1}^{t} - \delta \dot{\hat{\tau}}_{12}^{t}) \cdot   d_{32}^t\\
        - \delta\hat{\tau}_{12}^{t} + (1 + \delta \dot{\hat{\tau}}_{1}^{t}) \cdot d_{21}^t
    \end{bmatrix},\\
    \delta\dot{\hat{\tau}}_{ij}^{t}(x)&= \delta\dot{\hat{\tau}}_{i}^{t}(x) - \delta\dot{\hat{\tau}}_{j}^{t}(x).
\end{align}
Hence, the task is to solve six pseudoranges for nine variables (two relative SCET desynchronizations, six LTTs, one absolute SCET drift from TCB). This system is under-determined, so we must include additional measurements.\par
In \cref{sec:OrbitDetermination}, we show that ground-based measurements of the absolute SC positions and velocities, so-called orbit determinations (ODs), can be used to reduce the problem from six LTTs to three arm lengths. Furthermore, the telemetry data's onboard transmission and on-ground reception time tags can be used to estimate the absolute SCET desynchronizations from TCB (see \cref{sec:MocTimeCorrelation}). This MOC time correlation serves two purposes: (1) The estimation of the differential SCET desynchronizations via pseudorange disentanglement just enables relative synchronization of the three SCETs; for absolute synchronization to TCB, we additionally need information about the absolute desynchronization of one SCET from TCB, which is exactly what the MOC time correlation provides. (2) The time derivative of the MOC time correlation serves as an estimate for the absolute SCET drift with respect to TCB appearing in the pseudorange observation equations (see $\delta \dot{\hat{\tau}}_{1}^{t}$ in \cref{eq:AllPseudoranges}).

\section{Orbit Determination}\label{sec:OrbitDetermination}
For communication planning and SC navigation, the ESA tracks the absolute positions and velocities of the three LISA satellites. The ESA tracking stations send radio signals to the LISA SC, which are returned to Earth after a well-calibrated transponder delay. Range (propagation time) and Doppler measurements of these signals are processed in a square root information filter using an incremental window with processing arcs of typically several weeks duration \cite{Martens:OD}. This yields solutions for the absolute SC positions and velocities every few days. The errors of this orbit determination (OD) depend on the tracking direction: Conservative estimates by ESA state the uncertainties as  \SI{2}{\kilo\m} and \SI{4}{\milli\m\s\tothe{-1}}, \SI{10}{\kilo\m} and \SI{4}{\milli\m\s\tothe{-1}}, and \SI{50}{\kilo\m} and \SI{50}{\milli\m\s\tothe{-1}} in the along-track, radial-track (with respect to the sun), and cross-track directions, respectively. Due to the incremental window approach, we expect a smooth OD time series with correlated errors in subsequent ODs. Below, we show how the ODs can be used in the pseudorange disentanglement process despite their relatively high uncertainties.\par
The LTTs can be expressed in the BCRS in terms of the absolute positions and velocities of emitting and receiving SC \cite{Bayle:LisaOrbits,Chauvineau:LTT-Decomposition,Hees:LTT-Decomposition}:
\begin{align}
    d^t_{ij} &= \frac{1}{c}\: L^t_{ij} + \frac{1}{c^2}\: \vec{L}^t_{ij}\cdot \dot{\vec{x}}^t_{j} + O(c^{-3}),\label{eq:LightTravelTimeFromOrbitDetermination}\\
    \vec{L}^t_{ij} &= \vec{x}^t_{i} - \vec{x}^t_{j},\label{eq:VectorBetweenTwoSC}\\
    L^t_{ij}&= \sqrt{(x^t_i - x^t_j)^2 + (y^t_i - y^t_j)^2 + (z^t_i - z^t_j)^2},\label{eq:Armlengths}
\end{align}
where all quantities are evaluated at the TCB time of reception. $\vec{L}_{ij}$ denotes the vector pointing from the emitting SC at $\vec{x}_{j}$ to the receiving SC at $\vec{x}_{i}$, its absolute value $L_{ij}$ is the geometrical arm length. $\dot{\vec{x}}_j$ is the absolute velocity of the emitting SC. If we directly compose LTT estimates from the ODs according to \cref{eq:LightTravelTimeFromOrbitDetermination}, the OD errors couple into these estimates in the order of \SIrange[]{3}{70}{\kilo\m} depending on the link orientation (see \cref{sec:ODErrorCouplingArmlengths} for an analytical derivation and \cref{sec:2ndOrderLTC} for a numerical assessment). These errors far exceed the accuracy required for pseudorange disentanglement, prohibiting a direct usage of the OD results for this purpose.\par
\begin{figure}
    \begin{center}
        \includegraphics[width=0.5\textwidth]{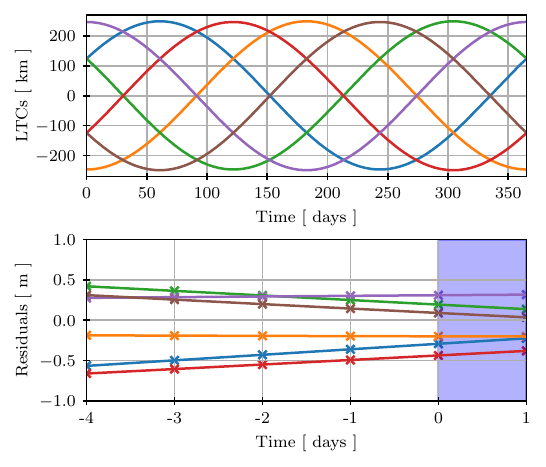}
    \end{center}
    \caption{Upper plot: A series of 365 LTCs for an ESA orbit file simulated with LISA orbits \cite{Bayle:LisaOrbits}. Lower plot: LTC estimation residuals for 6 ODs simulated with LISA ground tracking (see \cref{sec:Results}). Crosses indicate the actual ODs. Lines represent 5th-order spline interpolations. The time axis is aligned with the pseudorange measurements, which start at day 0 (see blue-shaded area and \cref{fig:MeasurementTimestamps}).}\label{fig:LTCsFromODs}
\end{figure}

We here express the LTTs in terms of the three geometrical arm lengths and six light time corrections (LTCs) denoted by $\Delta_{ij}$:
\begin{align}
    d^t_{ij} &= \frac{L^t_{ij}}{c} +\Delta^t_{ij},\\
    d^t_{ji} &= \frac{L^t_{ij}}{c} + \Delta^t_{ji},\\
    \Delta^t_{ij} &= \frac{1}{c^2}\: \vec{L}^t_{ij}\cdot \dot{\vec{x}}^t_{j} + O(c^{-3}).\label{eq:LightTimeCorrectionFromOrbitDetermination}
\end{align}
The LTCs can amount up to \SI{250}{\kilo\m} due to the heliocentric motion of the LISA satellites with about \SI{30}{\kilo\m\s\tothe{-1}} (see the upper part of \cref{fig:LTCsFromODs}).
At first glance, introducing LTCs does not seem to improve the situation since we now have to determine nine variables (three arm lengths and six LTCs). However, the OD uncertainty couples into the LTCs in the order of \SIrange[]{0.3}{0.7}{\m} (see \cref{sec:ODErrorCouplingLTC} for an analytical derivation, \cref{sec:2ndOrderLTC} for a numerical assessment, and the lower part of \cref{fig:LTCsFromODs} where we plot the LTC estimation errors of an exemplary OD time series). Thus, we can compose the LTCs from ODs according to \cref{eq:LightTimeCorrectionFromOrbitDetermination} at sufficient accuracy for our purpose. We neglect the LTC contributions of order $O(c^{-3})$. In \cref{sec:2ndOrderLTC}, we numerically assess that OD errors couple into these terms at \si{\milli\m} order, which is negligible.\par
In the pseudorange observation equations, we now treat the LTCs as parameters, which are fixed by the ODs. Thus, we reduce the number of variables from six LTTs to three arm lengths. The LTC error coupling into the pseudoranges can be computed to be
\begin{align}
    \delta R^t_{ij}(\Delta^t) = \frac{\partial R^t_{ij}}{\partial \Delta_{ij}^t} \cdot \delta \Delta^t = \left(1 + \delta \dot{\hat{\tau}}_{j}^{t}\right) \cdot \delta \Delta^t \approx \delta \Delta^t,
\end{align}
thus directly corresponding to the submeter LTC errors.

\section{MOC Time Correlation}\label{sec:MocTimeCorrelation}
\begin{figure*}
    \begin{center}
        \includegraphics[width=1\textwidth]{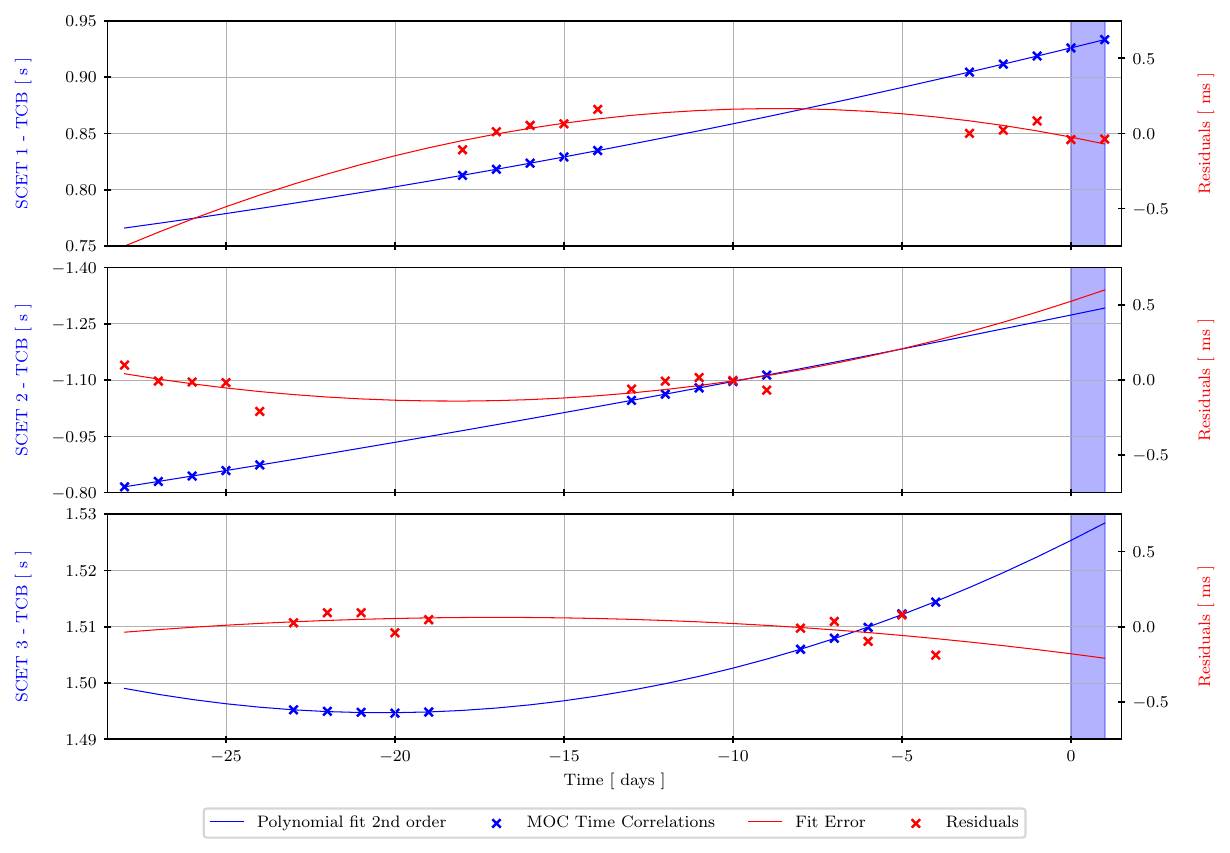}
    \end{center}
    \caption{A series of 30 MOC-TCs simulated with LISA Instrument \cite{Bayle:LisaInstrument} (see \cref{sec:Results}). Blue dots indicate the measurements, and red crosses the estimation residuals. Blue lines represent second-order polynomial fits, and red lines the associated errors. The MOC-TCs are separated by the target SCET: Upper, central, and lower plots show the MOC-TCs for SCET 1, SCET 2, and SCET 3, respectively illustrating the alternating LISA communication schedule. The time axes are aligned with the pseudorange measurements starting at day 0 (see blue-shaded area and \cref{fig:MeasurementTimestamps}).}\label{fig:MocTimeCorrelations}
\end{figure*}
The mission operation center (MOC) estimates the SCET desynchronization from TCB, a process to which we refer as MOC time correlation (MOC-TC) \cite{Klioner:MocTC}. Each telemetry packet is timestamped by the SCET of the transmitting SC at the event of transmission and by the TCB of reception on Earth. MOC computes the TCB corresponding to the transmission event to compare both time tags. This requires information about the SC-to-ground-station signal propagation delay that can be calculated from the ODs. It further involves proper modeling of relativistic corrections and atmospheric delays. The MOC-TC then provides a measurement of the difference between the SCET and the TCB times associated with the transmission event, i.e., a measurement of the SCET desynchronization from TCB:
\begin{align}
    \text{TC}^{t}_{i}(x) = \delta\hat{\tau}_{i}^{t}(x) + n_i.
\end{align}
The superscript $t$ indicates that the MOC-TC is timestamped in TCB. Its uncertainty $ n_i$ is estimated to be around \SI{0.1}{\milli\s}, which, among others, involves uncertainties of the SC-to-ground-station separation and uncertainties of the time tagging process on board.\par
Only one SC at a time communicates with Earth; hence, we only obtain MOC-TCs for one SCET at a time. We expect an alternating communication schedule in which the transmitting SC is changed every five days. This minimizes the number of required antenna repointings to once per 15 days.\footnote{Antenna repointings cause vibrations in the SC, thus potentially introducing noise. Therefore, their number should be kept at a minimum.} \Cref{fig:MocTimeCorrelations} shows a series of 30 MOC-TCs (blue dots) and the associated estimation residuals (red crosses) separated by the target SCET. The alternating LISA communication schedule can be seen here. The blue lines represent second-order polynomial fits; the red lines indicate the errors of these fits.\footnote{The time evolution of the SCET desynchronizations is clearly dominated by instrumental USO frequency offsets and linear drifts. Therefore, a second-order polynomial fit performs best.}\par
In the process of LISA clock synchronization, MOC-TCs serve two purposes: (1) They enable absolute synchronization to TCB. Their accuracy sets the limit for this absolute synchronization. For the relative SCET synchronization, pseudorange disentanglement yields 5 orders of magnitude higher accuracy in the order of \SI{1}{\nano\s} (see \cref{sec:Results}). (2) MOC-TCs are required for pseudorange disentanglement, where we have to estimate the absolute drift of one SCET from TCB (see $\delta\dot{\hat{\tau}}_{1}^{t}$ in \cref{eq:AllPseudoranges}). This drift can be deduced from the time derivative of a second-order polynomial fit of the MOC-TCs. For near-real-time pipelines, the fits of MOC-TCs for noncommunicating SC are less accurate, as they require extrapolation. Here, it is convenient to model the pseudoranges based on the absolute SCET drift of the currently communicating SC (SC 1 in the case of \cref{fig:MocTimeCorrelations}).\par
The MOC-TC error coupling into the pseudoranges can be computed to be
\begin{align}\label{eq:MOC-TC-error-coupling}
    \delta R^t_{ij}(\delta\dot{\hat{\tau}}_{1}^{t}) = \frac{\partial R^t_{ij}}{\partial\:\delta\dot{\hat{\tau}}_{1}^{t}} \cdot \delta \left(\delta\dot{\hat{\tau}}_{1}^{t}\right) = d_{ij}^t \cdot \delta \left(\delta\dot{\hat{\tau}}_{1}^{t}\right),
\end{align}
where $\delta \left(\delta\dot{\hat{\tau}}_{1}^{t}\right)$ denotes the uncertainty of the absolute SCET drift from TCB. We assess this coupling numerically in \cref{sec:MOC-TC-error-coupling}. For near-real-time pipelines, we observe an error coupling at submeter order. Posterior data improves the accuracy of the second-order polynomial fit so that the MOC-TC error coupling can be further reduced.

\section{Pseudorange disentanglement via iterative Kalman filtering}\label{sec:PseudorangeDisentanglement}
This section presents an algorithm for pseudorange disentanglement based on the iterative application of a modified Kalman filter (KF). Pseudorange disentanglement is a crucial step in the TDI baseline topology.
\begin{figure*}
    \begin{center}
        \includegraphics[width=1\textwidth]{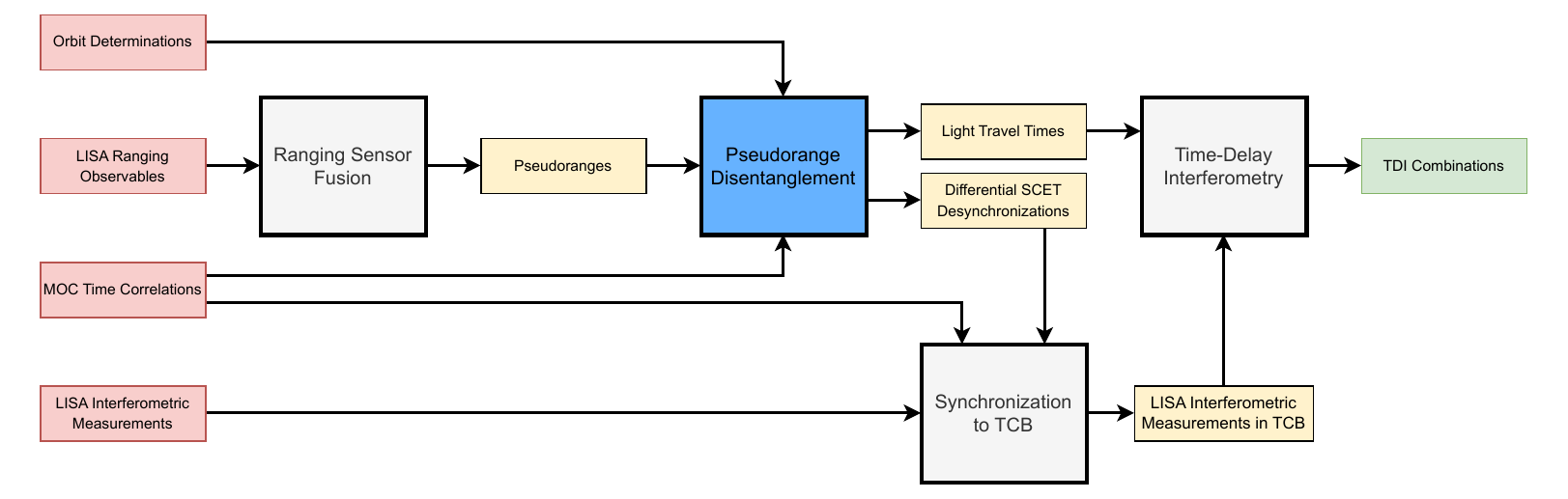}
    \end{center}
    \caption{Pseudorange disentanglement in the TDI baseline topology: The inputs are the ODs, the MOC-TCs, and the pseudoranges, which are delivered by a preceding ranging sensor fusion. The outputs are the light travel times used as delays in TDI and the differential SCET desynchronizations for the relative SCET synchronization. We neglect processing steps, which do not directly interfere with pseudorange disentanglement.}\label{fig:PseudorangeDisentanglementInTDIBaselineTopology}
\end{figure*}
It provides the required delays (light travel times) and the relative SCET desynchronizations, which, together with the MOC-TCs, facilitate the synchronization of the interferometric measurements to TCB (see \cref{fig:PseudorangeDisentanglementInTDIBaselineTopology}).\par
The here applied KF involves a linear system model and nonlinear measurement equations. We, therefore, refer to it as semi-extended KF. In contrast to previous studies \cite{Wang:KF-I,Wang:KF-II}, we choose a KF state vector that only includes relative quantities, which can be observed through the LISA interferometric measurements. Variables that can be fixed by ground-based observations are excluded from the state vector to ensure a well-defined system. We treat these variables as time-dependent external parameters. After establishing the semi-extended KF with external parameters in \cref{sec:SummaryKF}, we introduce the concrete system and observation models for the pseudorange disentanglement in \cref{sec:KalmanFilterPseudorangeDisentanglement}.\par
The standard KF assumes all state vector components and measurements to be given in some universal time frame. LISA data processing, however,  is concerned with four different time frames: The interferometric measurements are timestamped in the three SCETs, ODs and MOC-TCs in TCB. Our goal is to estimate the desynchronizations between these time frames. We address this difficulty by iterative Kalman filtering, which we elaborate in \cref{sec:IterativeKF}.

\subsection{The semi-extended Kalman filter with external parameters}\label{sec:SummaryKF}
Suppose we have a linear discrete-time system described by a state vector $x$, i.e., the time evolution of $x$ from $k$ to $k+1$ is given by the linear equation
\begin{align}\label{eq:LinearTimeDiscreteSystemPropagation}
    x_{k+1} &= F_k \cdot x_k + w_k.
\end{align}
$F_k$ governs the state vector's propagation in time and is referred to as the state transition matrix. The system dynamics are not perfectly known, a fact that is compensated by the process noise vector $w_k$. The associated noise process $\{w_k\}$ is assumed white and zero mean:
\begin{align}
    \text{E} \left[ w_{k} \cdot w^{\text{T}}_{l} \right] &= W_{k} \delta_{kl},\\
    w_{k} &\sim ( 0, W_k),
\end{align} 
where $W_k$ denotes the process noise covariance matrix and $\delta_{kl}$ the Kronecker delta.	The measurements are arranged in a measurement vector $y_{k}$, which can be related to the state vector and a set of time-dependent external parameters $p_k$ via nonlinear observation equations
\begin{align}\label{eq:NonlinearObservationEquation}
    y_{k} &= h_k(x_k,\:p_k,\:v_k).
\end{align}
$v_k$ denotes the measurement noise, the associated noise process $\{v_k\}$ is assumed white, zero mean, and uncorrelated with $\{w_k\}$:
\begin{align}
    \text{E} \left[ v_{k} \cdot v^{\text{T}}_{l} \right] &= V_{k} \delta_{kl},\\
    v_{k} &\sim ( 0, V_k),\\
    \text{E} \left[ w_{k} \cdot v^{\text{T}}_{l} \right] &= 0.
\end{align}
$V_k$ denotes the measurement noise covariance matrix.\par
Starting from some initial state vector estimate $\hat{x}^{+}_0$ with covariance matrix $\hat{P}^{+}_0$
\begin{align}
    \hat{x}^{+}_0 &= \text{E} \left[ x_0 \right],\\
    \hat{P}^{+}_0 &= \text{E} \left[ (x_0 - \hat{x}^{+}_0)\cdot (x_0 - \hat{x}^{+}_{0})^{\text{T}} \right],
\end{align}
 the Kalman filter propagates the system through time, incorporating all measurements taken on the way. Thus, it produces two kinds of state estimates: (1) State estimates after the propagation step (current measurements excluded) are referred to as \textit{a priori} and marked with a minus superscript: 
\begin{align}
    \hat{x}_k^{-} &= \text{E} \left[ x_{k}\vert y_1, y_2, ..., y_{k-1} \right],\\
    \hat{P}_k^{-} &= \text{E} \left[ (x_k - \hat{x}_k^{-})\cdot (x_k - \hat{x}_k^{-})^{\text{T}} \right].
\end{align}
(2) State estimates after the measurement update (current measurements included) are referred to as \textit{a posteriori} and marked with a plus superscript:
\begin{align}
    \hat{x}_k^{+} &= \text{E} \left[ x_{k}\vert y_1, y_2, ..., y_{k} \right],\\
    \hat{P}_k^{+} &= \text{E} \left[ (x_k - \hat{x}_k^{+})\cdot (x_k - \hat{x}_k^{+})^{\text{T}} \right].
\end{align}
The propagation from a posteriori state estimates to a priori state estimates at the subsequent time step is realized by (compare \cref{eq:LinearTimeDiscreteSystemPropagation})
\begin{align}
    \hat{x}_{k+1}^{-} &= F_k \cdot \hat{x}_{k}^{+},\label{eq:KalmanFilterPropagationStateVector}\\
    \hat{P}_{k+1}^{-} &= F_k\cdot \hat{P}_{k+1}^{+} \cdot F_k^{\text{T}} + W_k.\label{eq:KalmanFilterPropagationCovarianceMatrix}
\end{align}
To incorporate measurements at time $k$, we need to evaluate the external parameters at that time and linearize the observation model (\cref{eq:NonlinearObservationEquation}) around the a priori state estimate $\hat{x}_k^{-}$:
\begin{align}
    H_k = \frac{\partial  h_k(\hat{x}_k^{-},\:p_k,\:0)}{\partial x}\Bigg\vert_{\hat{x}_k^{-}}.\label{eq:KalmanFilterObservationModelLinearization}
\end{align}	
Then, $H_k$ is the observation matrix at time $k$. The measurement update from a priori to a posteriori state estimates is realized by
\begin{align}
    \hat{x}_k^{+} &= \hat{x}_k^{-} + K_k \cdot \left(y_k - h_k(\hat{x}_k^{-},\:p_k,\:0)\right),\label{eq:KalmanFilterUpdateStateVector}\\
    \hat{P}_k^{+}&=(\mathds{1} - K_k H_k) \cdot \hat{P}_{k}^{-},\label{eq:KalmanFilterUpdateCovarianceMartix}
\end{align}
where $K_k$ denotes the Kalman gain given by
\begin{align}
    K_k &= \hat{P}_k^- H_k^{\text{T}} \cdot (H_k P_k^- H_k^{\text{T}} + V_k)^{-1}.\label{eq:KalmanGain}
\end{align}
The semi-extended Kalman filter with external parameters can now be summarized as
\begin{enumerate}
    \item Initialize the state vector $\hat{x}^{+}_0$ and its covariance matrix $\hat{P}^{+}_0$.
    \item Compute the a priori estimates $\hat{x}_{k+1}^{-}$ and $\hat{P}_{k+1}^{-}$ from the a posteriori estimates of the preceding time step $\hat{x}_{k}^{+}$ and $\hat{P}_{k}^{+}$ via \cref{eq:KalmanFilterPropagationStateVector,eq:KalmanFilterPropagationCovarianceMatrix}.
    \item Evaluate the external parameters $p$ at time $k+1$.
    \item Linearize the observation model around the current a priori state estimate $\hat{x}_{k+1}^{-}$ according to \cref{eq:KalmanFilterObservationModelLinearization}.
    \item Compute the a posteriori estimates $\hat{x}_{k+1}^{+}$ and $\hat{P}_{k+1}^{+}$ combining the current measurements $y_k$ with the a priori estimates $\hat{x}_{k+1}^{-}$ and $\hat{P}_{k+1}^{-}$ via \cref{eq:KalmanFilterUpdateStateVector,eq:KalmanFilterUpdateCovarianceMartix,eq:KalmanGain}.
    \item $k \leftarrow k+1$ and restart from step 2.
\end{enumerate}
After incorporating all measurements, we apply the RTS smoothing algorithm \cite{Rauch:Smoothing}.

\subsection{The Kalman filter for pseudorange disentanglement}\label{sec:KalmanFilterPseudorangeDisentanglement}
\subsubsection{The state vector and how to initialize it}
The measurements for pseudorange disentanglement are the six pseudoranges obtained from a preceding ranging sensor fusion \cite{Reinhardt:RangingSensorFusion}. For Kalman filtering, they must be expressed in terms of a state vector: Considering the pseudorange decomposition in \cref{eq:AllPseudoranges}, we would naively conclude that this state vector needs to contain the six LTTs, the two differential SCET desynchronizations, and one absolute SCET drift with respect to TCB. However, such a system is under-determined (see \cref{sec:ClockSynchronizationViaPseudorangeDisentanglement}). It is convenient to express the LTTs in terms of three arm lengths and six LTCs (see \cref{sec:OrbitDetermination}). While the arm lengths are included in the state vector, the LTCs are fixed by the ODs and treated as time-dependent external parameters. Similarly, we proceed with the absolute SCET drift from TCB: We deduce it from the MOC-TCs and treat it as an external parameter (see \cref{sec:MocTimeCorrelation}). Hence, the state vector $x$ and the external parameters $p$ are given by
\begin{align}
    x &= x_{\text{orbit}} \oplus x_{\text{clock}} \in \mathbb{R}^{15},\\
    x_{\text{orbit}} &= \left[ L_{12},L_{23},L_{31},\:\dot{L}_{12},\dot{L}_{23},\dot{L}_{31},\:\ddot{L}_{12},\ddot{L}_{23},\ddot{L}_{31}\right],\\
    x_{\text{clock}} &= \left[ \delta \hat{\tau}_{12},\delta \hat{\tau}_{13},\:\delta \dot{\hat{\tau}}_{12},\delta \dot{\hat{\tau}}_{13},\:\delta \ddot{\hat{\tau}}_{12},\delta \ddot{\hat{\tau}}_{13}\right],\\
    p &= \left[\Delta_{ij},\:\dot{\Delta}_{ij},\:\delta \hat{\tau}_1^t, \:\delta \dot{\hat{\tau}}_1^t\right],\:i\neq j \in \{1,\:2,\:3\}.
\end{align}
Here clock 1 is chosen as reference. Other choices are possible, e.g., the mean of all clocks. To avoid discrepancies between the dynamic range of $x_{\text{orbit}}$ and $x_{\text{clock}}$, we consider the arm lengths in seconds, i.e., in $x_{\text{orbit}}$ we actually mean $L_{ij} / \text{c}$, etc. Thus, the orders of magnitude are comparable.\par
The orbital part of the state vector can be initialized using orbit determinations. We compose estimates for the arm lengths via \cref{eq:Armlengths}. The first and second-order time derivatives can be set up according to
\begin{align}
    \dot{L}_{ij} &= \frac{\vec{L}_{ij}\cdot \dot{\vec{L}}_{ij}}{L_{ij}},\label{eq:ArmlengthDerivative}\\
    \ddot{L}_{ij} &= \frac{\left\vert\dot{\vec{L}}_{ij}\right\vert^2}{L_{ij}} + \frac{\dot{\vec{L}}_{ij}\cdot \ddot{\vec{L}}_{ij}}{L_{ij}} - \frac{\left(\vec{L}_{ij}\cdot \dot{\vec{L}}_{ij}\right)^2}{L^3_{ij}},
\end{align}
where $\dot{\vec{L}}_{ij}$ and $\ddot{\vec{L}}_{ij}$ are the time derivatives of \cref{eq:VectorBetweenTwoSC}. $x_{\text{clock}}$ could be initialized using MOC-TCs of all three SC. To demonstrate the robustness of the algorithm, we restrict it to the MOC-TCs of just one SC: We initialize the clock part of the state vector with zeros and use the MOC-TCs of the currently communicating SC to set up the external parameter $\delta \dot{\hat{\tau}}_1^t$.\footnote{During operation the initialization is just required at the commissioning and after interruptions. Otherwise, it is continuously available.}\par 
To set up the covariance matrix associated with the initial state vector we consider the coupling of OD uncertainties into the estimates of $L_{ij}$, $\dot{L}_{ij}$, and $\ddot{L}_{ij}$ (see \cref{sec:ODErrorCouplingArmlengths}), and on-ground calibrations of space-qualified ultrastable oscillators \cite{Asmar:USO,Hartwig:TDIwoSync}:
\begin{align}
    \hat{P}^{+}_{\text{orbit},\:0} &= \text{diag}\Big[
    \SI{2e-4}{\s}\: \ldots \: \num{e-9}\: \ldots \:\SI{e-15}{\s\tothe{-1}}
    \Big]^2,\nonumber\\
    \hat{P}^{+}_{\text{clock},\:0} &= \text{diag}\Big[
    \SI{1}{\s} \:\ldots \: \num{e-7}\: \ldots \:\SI{e-14}{\s\tothe{-1}}
    \Big]^2.
\end{align}
In reality, $\hat{P}^{+}_0$ will not be a diagonal matrix. The estimates for the different arm lengths are obtained from the same OD data, hence we expect strong correlations between them.
\subsubsection{The system model}
The state transition matrix $F$ is a block-diagonal matrix, i.e., the orbit and clock models can be separated:
\begin{align}
    F_k &= \begin{bmatrix}
    \textbf{F}_{\text{orbit}} & \textbf{0}_{\:9\times 6}\\
    \textbf{0}_{\:6\times 9} & \textbf{F}_{\text{clock}}
    \end{bmatrix},\\
    \textbf{F}_{\text{orbit}} &= \begin{bmatrix}
    1 & 0 & 0 & \Delta t & 0 & 0 & \frac{\Delta t^2}{2} & 0 & 0\\
    0 & 1 & 0 & 0 & \Delta t & 0 &0 & \frac{\Delta t^2}{2} & 0\\
    0 & 0 & 1 & 0 & 0 & \Delta t & 0 & 0 & \frac{\Delta t^2}{2}\\
    0 & 0 & 0 & 1 & 0 & 0 & \Delta t & 0 & 0\\
    0 & 0 & 0 & 0 & 1 & 0 & 0 & \Delta t & 0\\
    0 & 0 & 0 & 0 & 0 & 1 & 0 & 0 & \Delta t\\
    0 &	0 & 0 & 0 & 0 & 0 & 1 & 0 & 0\\
    0 &	0 & 0 & 0 & 0 & 0 & 0 & 1 & 0\\
    0 &	0 & 0 & 0 & 0 & 0 & 0 & 0 & 1
    \end{bmatrix},\\
    \textbf{F}_{\text{clock}} &= \begin{bmatrix}
    1 & 0 & \Delta t & 0 & \frac{\Delta t^2}{2} & 0\\
    0 & 1 & 0 & \Delta t & 0 & \frac{\Delta t^2}{2}\\
    0 & 0 & 1 & 0 & \Delta t & 0\\
    0 & 0 & 0 & 1 & 0 & \Delta t\\
    0 & 0 & 0 & 0 & 1 & 0\\
    0 & 0 & 0 & 0 & 0 & 1
    \end{bmatrix},
\end{align}
where $\Delta t $ denotes the sampling time. We model the process noise covariance matrix as\footnote{These values are obtained heuristically. A follow-up study could investigate a more sophisticated process-noise modeling.}
\begin{align}	
    &W = \text{diag}\Big[0, 0, 0, 0, 0, \SI{e-13}{\per \s},  \SI{e-13}{\per \s}, \SI{e-13}{\per \s},\nonumber\\
    &0, 0, 0, 0, \SI{e-13}{\per \s},  \SI{e-13}{\per \s}\Big]^2.
\end{align}

\subsubsection{The observation model}
We now need to express the measurements in terms of the state vector components and the external parameters. Essentially, this is done in \cref{eq:AllPseudoranges}; we just need to replace the LTTs by arm lengths and LTCs:
\begin{align}\label{eq:ObservationModel}
    &h(x_k,\:p_k,\:0) =\nonumber \\
    &\begin{bmatrix}
    \delta\hat{\tau}_{12}^{t} + (1 + \delta \dot{\hat{\tau}}_{1}^{t} - \delta \dot{\hat{\tau}}_{12}^{t}) \cdot (L_{12}^t + \Delta_{12}^t)\\
    \delta\hat{\tau}_{13}^{t} - \delta\hat{\tau}_{12}^{t} + (1 + \delta \dot{\hat{\tau}}_{1}^{t} - \delta \dot{\hat{\tau}}_{13}^{t}) \cdot (L_{23}^t + \Delta_{23}^t)\\
    - \delta\hat{\tau}_{13}^{t} + (1 + \delta \dot{\hat{\tau}}_{1}^{t}) \cdot (L_{31}^t + \Delta_{31}^t)\\
    \delta\hat{\tau}_{13}^{t} + (1 + \delta \dot{\hat{\tau}}_{1}^{t} - \delta \dot{\hat{\tau}}_{13}^{t}) \cdot (L_{31}^t + \Delta_{13}^t)\\
    \delta\hat{\tau}_{12}^{t} - \delta\hat{\tau}_{13}^{t} + (1 + \delta \dot{\hat{\tau}}_{1}^{t} - \delta \dot{\hat{\tau}}_{12}^{t}) \cdot  (L_{23}^t + \Delta_{32}^t)\\
    - \delta\hat{\tau}_{12}^{t} + (1 + \delta \dot{\hat{\tau}}_{1}^{t}) \cdot (L_{12}^t + \Delta_{21}^t)
    \end{bmatrix}_k.
\end{align}
We model the measurement noise covariance matrix as
\begin{align}
    V = (\SI{1e-9}{\s})^2 \cdot \mathds{1}_{6}.
\end{align}
From the ranging sensor fusion, we actually obtain pseudorange RMS values in the order of \SI{1e-11}{\s}. Here, we choose increased measurement noise levels to put more emphasis on the system model.

\subsection{Iterative Kalman filtering}\label{sec:IterativeKF}
While the standard KF assumes all measurements and all state vector components to be given in some universal time frame, pseudorange disentanglement is concerned with four different time frames: The six pseudoranges, recorded according to the respective SCETs, need to be combined with ground-based observations, which are given in TCB. We approach this difficulty by iterative Kalman filtering: In the first iteration, we assume all pseudoranges to be given in TCB. We process them in the above-described KF. Thus, we get first estimates for the differential SCET desynchronizations $\delta\hat{\tau}_{12,\:\text{kf}}^{t},\:\delta\hat{\tau}_{13,\:\text{kf}}^{t}$. We combine them with the MOC-TCs to time shift the pseudoranges from the three SCETs to TCB:
\begin{align}
    \delta\hat{\tau}_{1,\:\text{shift}}^{t}(x) &= \text{TC}^{t}_{1}(x),\\
    \delta\hat{\tau}_{2,\:\text{shift}}^{t}(x)  &= \text{TC}^{t}_{1}(x)  - \delta\hat{\tau}_{12,\:\text{kf}}^{t}(x),\\
    \delta\hat{\tau}_{3,\:\text{shift}}^{t}(x)  &= \text{TC}^{t}_{1}(x)  - \delta\hat{\tau}_{13,\:\text{kf}}^{t}(x),
\end{align}
The time shifting splits the pseudoranges according to the SC they are taken on: $R_{12}$ and $R_{13}$, which are measured on SC 1, experience different time shifts than $R_{23}$ and $R_{21}$, which are taken on SC 2, and $R_{31}$ and $R_{32}$, which are taken on SC 3. We interpolate the time-shifted pseudoranges using 5'th order Lagrange interpolation and resample them uniformly. We then execute the KF again, using the time-shifted interpolated pseudoranges. In the subsequent section, we show that the results converge after the second iteration.

\section{Results}\label{sec:Results}

\subsection{Data simulation}\label{sec:Data-simulation}
\begin{figure}
    \begin{center}
        \includegraphics[width=0.5\textwidth]{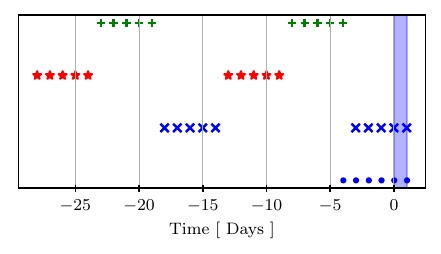}
    \end{center}
    \caption{Measurement timestamps: Crosses, stars, and plus marks indicate the MOC-TCs for SCET 1, SCET 2, and SCET 3, respectively. Dots represent the orbit determinations. The range of the pseudoranges is shaded in blue.}\label{fig:MeasurementTimestamps}
\end{figure}
In this section, we assess the performance of the pseudorange disentanglement algorithm established in \cref{sec:PseudorangeDisentanglement}. We consider one day of telemetry data simulated with LISA Instrument \cite{Bayle:LisaInstrument} based on an orbit file provided by ESA \cite{Bayle:LisaOrbits}. The blue area in \cref{fig:MeasurementTimestamps} indicates the range of the telemetry data. We simulate the LTTs up to and including the $c^{-3}$ terms as specified in \cref{sec:2ndOrderLTC}. For the SCET desynchronizations from TCB, we include deviations of the three SC proper times from TCB, which are simulated according to \cite{Bayle:LisaOrbits}
\begin{align}\label{eq:TPSwrtTCB}
    \delta \dot{\tau}_i^t(x) = -\frac{1}{2} \left( \frac{r_s}{\abs{\vec{x}_i(x)}} + \frac{\abs{\vec{v}_i(x)}^2}{c^2}\right). 
\end{align}
Here $x$ is used as a generic function argument, $r_s$ denotes the Schwarzschild radius of the sun, and $\vec{x}_i$ and $\vec{v}_i$ are the SC positions and velocities. We further include instrumental deviations of the three SCETs from these proper times, which are simulated according to the clock model described in \cite{Hartwig:TDIwoSync}:
\begin{align}\label{eq:Clock-model}
    \delta \hat{\tau}_{i}^{\tau_i}(x) = \delta \hat{\tau}_{i,\:0} + y_i \: x + \frac{\dot{y}_i}{2} \: x^2 + \frac{\ddot{y}_i}{3} \: x^3 + \int_{x_0}^{x} \text{d} \tilde{x} \: y_{i}^{\epsilon}(\tilde{x}).
\end{align}
The initial SCET offsets $\delta \hat{\tau}_{i,\:0}$ are set to \SI{1.6}{\s}, \SI{-0.9}{\s}, and \SI{0.4}{\s} for SC 1, 2, and 3, respectively. We consider USO frequency offsets $y_i \sim 10^{-7}$ (in fractional frequency deviations), linear USO frequency drifts $\dot{y}_i\sim10^{-14}\:\si{\s\tothe{-1}}$, and quadratic drifts $\ddot{y}_i\sim10^{-23}\:\si{\s\tothe{-2}}$. We include stochastic clock noise $y_{i}^{\epsilon}$ with the amplitude spectral density (ASD)
\begin{align}
    \sqrt{S_{y^{\epsilon}}(f)} = 6.32 \times 10^{-14} \si{\hertz\tothe{-0.5}} \left(\frac{f}{\si{\hertz}}\right)^{-0.5}.
\end{align}
Furthermore, we consider laser frequency, ranging, modulation, test-mass, and readout noise as specified in \cite{Bayle:LISA-Simulation}.\par
On this telemetry data set, we apply the ranging sensor fusion methods described in \cite{Reinhardt:RangingSensorFusion}. The output pseudoranges and their time derivatives are the input to the pseudorange disentanglement processing element (see \cref{fig:PseudorangeDisentanglementInTDIBaselineTopology}). Their precision is limited by left-handed modulation noise with the ASD
\begin{align}
    \sqrt{S_{M}(f)} &= 8.3 \times 10^{-15}\si{\s\hertz\tothe{-0.5}}\left(\frac{f}{\si{Hz}}\right)^{-2/3}.
\end{align}\par
The ODs are simulated with LISA ground tracking. For each SC we draw errors for the absolute position and velocity measurements in the three directions (along, radial, and cross-track) from Gaussian distributions. The conservative estimates for the OD uncertainties specified in \cref{sec:OrbitDetermination} are used as standard deviations. We linearly propagate these errors in time, which yields a smooth time series with correlated errors in subsequent ODs (see the lower plot in \cref{fig:LTCsFromODs}). We expect this to be close to reality considering the incremental window approach that is applied in the processing of the range and Doppler measurements (see \cref{sec:OrbitDetermination}). We simulate six ODs with a sampling time of one day (see blue dots in \cref{fig:MeasurementTimestamps}).\par
The MOC-TCs are simulated in LISA Instrument \cite{Bayle:LisaInstrument}. We expect their errors to be dominated by onboard time tagging inaccuracies, which we model as white noise with an RMS amplitude of \SI{0.1}{\milli\s}. We simulate a series of 30 MOC-TCs in the alternating LISA communication schedule (see crosses, stars, and plus marks in \cref{fig:MeasurementTimestamps}). We just use the MOC-TCs associated with SCET 1 since SC 1 is the currently communicating one in this example.\par
\begin{figure}
    \begin{center}
        \includegraphics[width=0.5\textwidth]{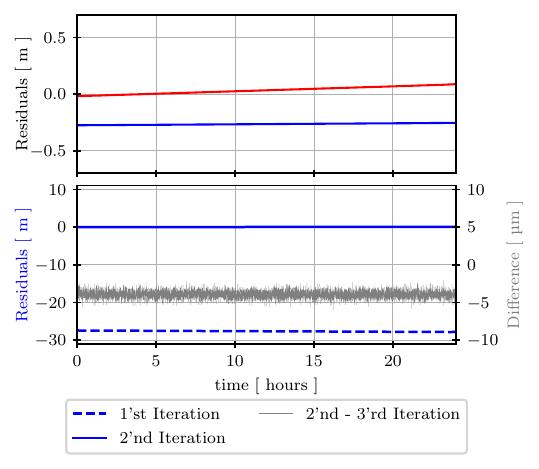}
    \end{center}
    \caption{Estimated differential SCET desynchronizations for one exemplary realization of the ground-based measurements. We plot the residuals, i.e., the true values are subtracted. Upper plot: $\delta \hat{\tau}_{12}$ (blue) and $\delta \hat{\tau}_{13}$ (red) after two KF iterations. Lower plot: $\delta \hat{\tau}_{12}$ after one iteration (dashed blue) and after two iterations (solid blue). The difference between the second and the third iteration is plotted in black (see right y-axis).}\label{fig:EstimatedDifferentialScetOffsetOneRealization}
\end{figure}
The LTCs and the SCET drift with respect to TCB appearing in the pseudorange equations are determined by only a few ODs and MOC-TCs. Hence, the pseudorange disentanglement accuracy depends on the particular noise realization of these ground-based measurements. To assess the performance of our algorithm, we conduct a Monte Carlo simulation combining the pseudorange data set obtained from the ranging sensor fusion with 1000 realizations of ODs and MOC-TCs.

\subsection{Clock synchronization}
\begin{figure}
    \begin{center}
        \includegraphics[width=0.5\textwidth]{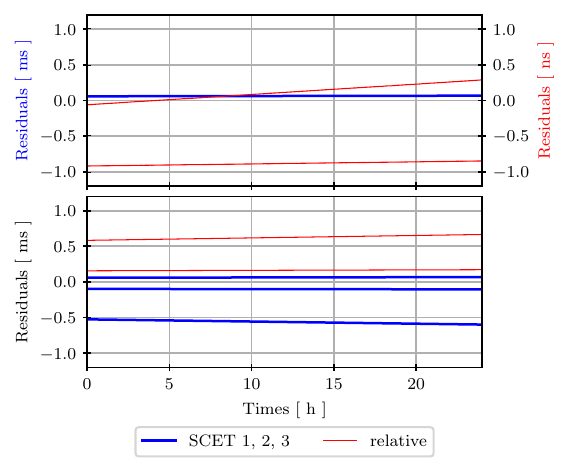}
    \end{center}
    \caption{Absolute SCET synchronization to TCB. Upper plot: We use the MOC-TCs for SC 1 and disentangle the pseudoranges. Blue: Residual absolute SCET desynchronization from TCB (all three SC). Red: Residual differential SCET desynchronizations (see right y-axis). Lower plot: We do not disentangle the pseudoranges and use MOC-TCs for all three SC (see \cref{fig:MocTimeCorrelations}).}\label{fig:EstimatedAbsoluteScetOffsetOneRealization}
\end{figure}
\begin{figure}
    \begin{center}
        \includegraphics[width=0.5\textwidth]{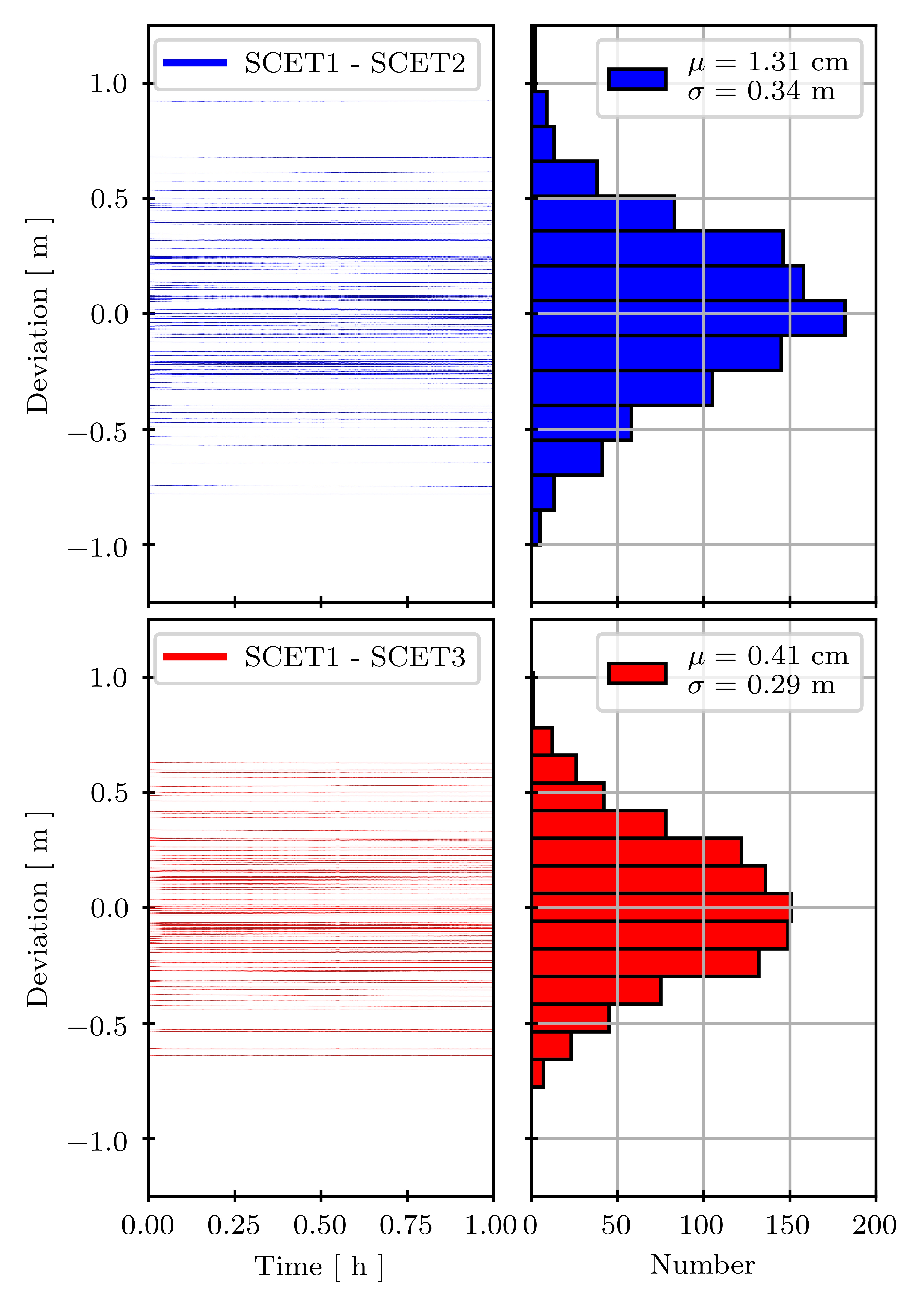}
    \end{center}
    \caption{Estimated differential SCET desynchronizations for one hour of pseudoranges with 1000 different realizations of ODs and MOC-TCs. Left plots: Time series of the residuals for every tenth realization. Right plots: Histogram plot of the average residual desynchronizations.}\label{fig:MonteCarloSimulationDifferentialSCETOffsets}
\end{figure}
Pseudorange disentanglement yields estimates for the differential SCET desynchronizations. The upper plot in \cref{fig:EstimatedDifferentialScetOffsetOneRealization} shows the residual differential SCET desynchronizations after two KF iterations for 24 hours of pseudorange data with one exemplary realization of the ODs and the MOC-TCs. Residual means that we subtract the true values (obtained as validation support from LISA Instrument) from the estimates. The differential SCET desynchronizations can be estimated with submeter accuracy. The residual errors drift with a few \si{\centi\m} per day and are caused by time-dependent OD and MOC-TC errors (see \cref{fig:LTCsFromODs,fig:MocTimeCorrelations}). The lower plot shows the different iterations for $\delta \hat{\tau}_{12}$: The second iteration increases the estimation accuracy by almost 2 orders of magnitude. The estimates converge after two iterations: The black line shows the difference between the second and the third iteration, which reaches the numerical noise floor.\par
The MOC-TCs limit the overall SCET synchronization to TCB. The upper plot in \cref{fig:EstimatedAbsoluteScetOffsetOneRealization} shows the performance of our algorithm: We reach a synchronization accuracy of better than \SI{0.1}{\milli\s} for all three SCETs (blue line). The lines cannot be distinguished because of the achieved nanosecond accuracy for the relative SCET synchronization (see red lines and right y-axis).\par
For comparison, we synchronize the SCETs without pseudorange disentanglement using MOC-TCs of all three SC (see lower plot in \cref{fig:EstimatedAbsoluteScetOffsetOneRealization}). For near-real-time pipelines, this procedure requires extrapolating the MOC-TCs associated with the two noncommunicating SC. Here, the reached accuracy is slightly lower. For the currently communicating SC 1 the results do not change. The true strength of the pseudorange disentanglement procedure lies in the relative SCET synchronization, where it outperforms the method based on MOC-TCs by about 6 orders of magnitude (compare the red lines in the upper and the lower plot of \cref{fig:EstimatedAbsoluteScetOffsetOneRealization}).\par
The pseudorange disentanglement accuracy is limited by the LTCs and the absolute drift of SCET 1 from TCB, which are fixed by only a few ODs and MOC-TCs. To assess the performance of our algorithm, we conduct a Monte Carlo simulation with 1000 realizations of these ground-based measurements. \Cref{fig:MonteCarloSimulationDifferentialSCETOffsets} shows the results for the relative SCET synchronization. On the left-hand side, we plot the time series of the residual desynchronizations. The right-hand side displays the average residual desynchronizations in histogram form: We obtain the standard deviations $\sigma = \SI{0.34}{\m}$ for $\delta \hat{\tau}_{12}$ and $\sigma = \SI{0.29}{\m}$ for $\delta \hat{\tau}_{13}$. This confirms that pseudorange disentanglement can deliver relative SCET synchronization with submeter accuracy.

\subsection{Light travel time estimation}
\begin{figure}
    \begin{center}
        \includegraphics[width=0.5\textwidth]{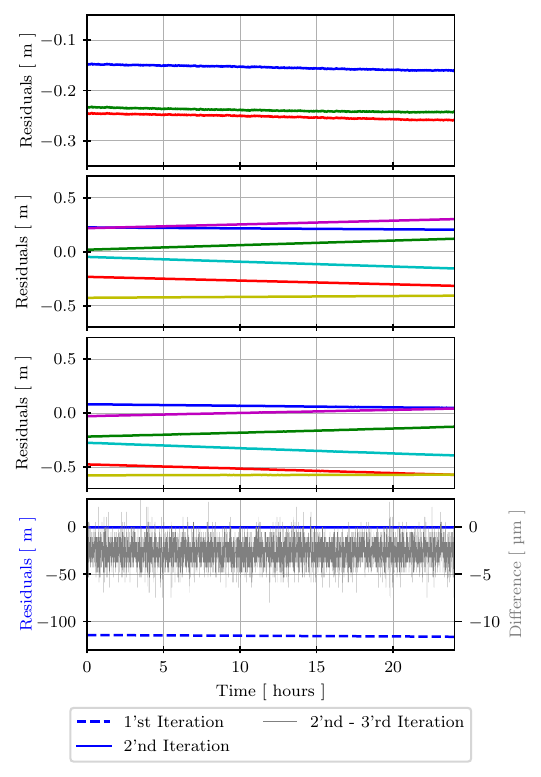}
    \end{center}
    \caption{Estimated arm lengths for one exemplary realization of ODs and MOC-TCs. Upper plot: Arm length estimates after 2 KF iterations. Second plot: LTCs composed of ODs. Third plot: LTT estimates. Lower plot: Estimated arm length 12 after 1 iteration (blue dashed) and after 2 iterations (blue solid). The difference between the second and the third iteration is plotted in black.}\label{fig:ArmlengthsEstimatesOneRealization}
\end{figure}
\begin{figure}
    \begin{center}
        \includegraphics[width=0.5\textwidth]{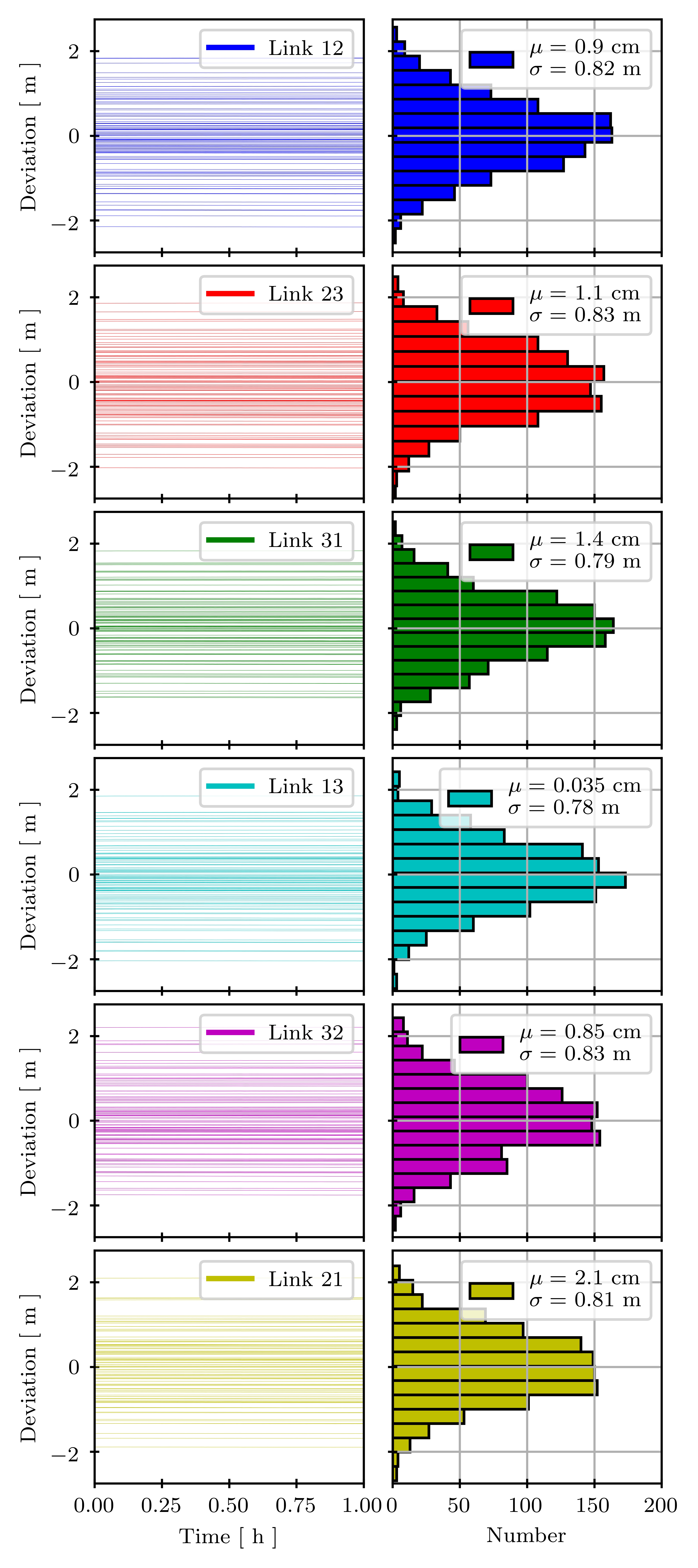}
    \end{center}
    \caption{LTT estimation for one hour of pseudoranges with 1000 different realizations of the ground-based measurements. Left plots: Residual LTT time series of every tenth realization. Right plots: Histogram plot of the average errors.}\label{fig:MonteCarloSimulationArmlengths}
\end{figure}
Pseudorange disentanglement yields estimates for the LTTs. As explained in \cref{sec:PseudorangeDisentanglement}, we consider the arm lengths in the KF state vector and treat the LTCs as external parameters. Combined, they form LTT estimates. The upper plot in \cref{fig:ArmlengthsEstimatesOneRealization} shows the arm length estimates after two KF iterations for 24 hours of pseudorange data with one exemplary realization of the ground-based measurements. The second plot shows the six LTCs (from ODs). In the third plot, we present the estimates for the six LTTs (essentially the sum of the upper two plots). After two KF iterations, the LTTs can be estimated with submeter accuracy. The lower plot shows the different iterations for the arm length 12: The second iteration increases the estimation accuracy by 2 orders of magnitude. Again, the estimates converge after two iterations.\par
Fig. \ref{fig:MonteCarloSimulationArmlengths} shows the results of the Monte Carlo simulation for the LTT estimation: The plots on the left-hand side show the time series of the LTT estimation residuals for the different realizations of the ground-based measurements. On the right-hand side, we plot their averages in histogram form. The standard deviations for the different links are between \SI{0.78}{\centi\m} and \SI{0.83}{\centi\m}. Hence, pseudorange disentanglement can deliver LTT estimates with submeter accuracy.\par
Note that these results are based on conservative estimates for the OD and MOC-TC performance. Furthermore, we expect the accuracy of the external parameters to increase with the number of available ODs and MOC-TCs, leading to a better pseudorange disentanglement performance on the long run. Finally, the estimation errors for the LTTs and the differential SCET desynchronizations are correlated (see \cref{sec:Correlations}). While the individual estimation errors are in the order of \SI{80}{\centi\m} in the case of the LTTs and \SI{30}{\centi\m} for the differential SCET desynchronizations, the combined errors are in the order of a few \si{\centi\m} (see \cref{fig:Correlations}). When executing the TDI baseline topology, TDI will be limited by the combined errors, not by the individual ones.

\section{Conclusion and Outlook}\label{sec:Conclusion}
To suppress laser frequency noise below the secondary noise levels, the TDI baseline topology requires LTTs and relative SCET synchronization with an accuracy of about \SI{10}{\m} \cite{Staab:TDI-Accuracy}. This paper presents an algorithm for pseudorange disentanglement that delivers LTTs and differential SCET desynchronizations with submeter accuracy. Due to correlations, the combined error of these estimates is in the order of a few \si{\centi\m}. This combined error is what poses the limit to TDI. Consequently, our processing element is ready to be interfaced with the TDI baseline topology, which we intend to do in a follow-up investigation.\par
We drop the simplifications made in previous investigations and apply a realistic LISA simulation: We consider different LTTs for counterpropagating laser links; we properly simulate the emission and reception times in the pseudorange modeling; we consider the SCET desynchronizations in the measurement timestamps; we use ESA orbits and apply relativistic modeling of the SC proper times \cite{Bayle:LisaOrbits}. Apart from that, we build our KF state vector upon relative interspacecraft variables, which the LISA interferometric measurements can actually observe. We exclude quantities that can be fixed by ODs and MOC-TCs to keep the number of measurements higher than the number of unknown state vector components. By restricting the state vector to relative positions (arm lengths), we avoid substantial nonlinearities in our observation equations, which arise in expressions of LTTs in terms of absolute SC positions (see \cref{sec:2ndOrderLTC}) \cite{Wang:KF-I}. Furthermore, this keeps the dynamic range of our state vector at acceptable orders of magnitude.\par
The ODs and MOC-TCs limit our pseudorange disentanglement performance. Previous studies did not have to include these ground-based measurements due to the above mentioned simplifications in the modeling and could, thus, yield a higher accuracy for the pseudorange disentanglement. In this paper, we use conservative estimates for the OD and MOC-TC uncertainties and show that the performance of our algorithm still fulfills the TDI requirements. A follow-up study should investigate pseudorange disentanglement for different uncertainties of the ground-based measurements. Such an investigation could set performance requirements for the ODs and the MOC-TCs. Moreover, the white noise model for the MOC-TCs is based on vague assumptions. We need a realistic model for the MOC-TCs to investigate their influence on pseudorange disentanglement accurately.\par
It must be emphasized that the here applied KF is nonstandard as it combines measurements recorded according to different time frames. We resolve this with an iterative procedure: After each iteration, we combine the estimated relative SCET desynchronizations with the MOC-TCs to timeshift the measurements to TCB, interpolate them, and resample them uniformly. The disadvantage of this procedure is that potential measurement interpolation errors could couple into the subsequent iterations. Measurement interpolation could be avoided using a sequential KF that incorporates the timeshifted measurements on an irregular time grid. A follow-up study should compare both methods.\par
While pseudorange disentanglement is an indisputable step in the TDI baseline topology, the alternative TDI topology is performed before clock synchronization and uses the pseudorange estimates obtained from a ranging sensor fusion \cite{Hartwig:TDIwoSync,Reinhardt:RangingSensorFusion}. The final TDI combinations, timestamped in the respective SCETs here, must still be synchronized with TCB. The accuracy requirement for this post-TDI synchronization is usually stated as \SI{0.5}{\milli\s}. In principle, this accuracy could be reached using MOC-TCs of all three SC. However, it might be advantageous to apply pseudorange disentanglement also here: Clock synchronization via pseudorange disentanglement is more robust as it only relies on the MOC-TCs of the currently communicating SC. Thus, it avoids extrapolation of MOC-TCs in near-real-time pipelines. Note that this requires updating the KF models every five days due to the alternating LISA communication schedule.\par
The true advantage of the pseudorange disentanglement algorithm lies in the relative SCET synchronization, where it outperforms the procedure relying on MOC-TCs by 6 orders of magnitude. The alternative TDI topology can also benefit from optimal relative SCET synchronization: (1) ESA sends commands to LISA with execution timestamps. A command is executed when the respective SCET matches the timestamp given on Earth. To simultaneously execute commands on the three LISA satellites, ESA needs to know the differential SCET offsets as accurately as possible to compensate for them when assigning the execution timestamps on Earth. (2) Relative SCET synchronization helps to study the effect of events like solar flares, which hit the three SC almost simultaneously.\par
It could be argued whether it is advantageous to disentangle the pseudoranges in a LISA-centered reference frame. We still have to consider LTCs. However, in a LISA-centered frame, they are about 3 orders of magnitude smaller and much more stable since they are mainly caused by the Sagnac effect associated with the cartwheel rotation. Hence, the uncertainties of LTCs are expected to be lower in a LISA-centered frame than in the BCRS. This could lead to more accurate LTTs, which might be particularly interesting from the perspective of TDI. However, LISA data analysis is performed in the BCRS. After relative SCET synchronization in a LISA-centered frame, we still need to perform a Poincaré transformation to the BCRS, and this is where the OD uncertainties come into play (Peter Wolf, personal communication, August 2022). A follow-up investigation could compare pseudorange disentanglement in both frames.

\appendix
\section*{Acknowledgements}
The authors thank the on-ground processing expert group for fruitful discussions. The authors thank the LISA simulation expert group for all simulation-related activities. The authors thank Jean-Baptiste Bayle, Aurélien Hees, and Waldemar Martens for their assistance in the search for some very mean bugs. The authors acknowledge support by the German Aerospace Center (DLR) with funds from the Federal Ministry of Economics and Technology (BMWi) according to a decision of the German Federal Parliament (Grant No. 50OQ2301, based on Grants No. 50OQ0601, No. 50OQ1301, No. 50OQ1801). Furthermore, this work was supported by the LEGACY cooperation on low-frequency gravitational wave astronomy (M.IF.A.QOP18098) and by the Bundesministerium für Wirtschaft und Klimaschutz based on a resolution of the German Bundestag (Project Ref. Number 50 OQ 1801). J. N. R. acknowledges the funding by the Deutsche Forschungsgemeinschaft (DFG, German Research Foundation) under Germany's Excellence Strategy within the Cluster of Excellence PhoenixD (EXC 2122, Project ID 390833453). J. N. R. further acknowledges the support by the IMPRS on gravitational-wave astronomy at the Max-Planck Institute for Gravitational Physics in Hannover, Germany. 

\section{OD error coupling}\label{sec:ODErrorCoupling}
In this section, we analytically compute the coupling of OD errors into the estimates of arm lengths and light time corrections. We assume all absolute position and velocity measurements to be uncorrelated so that we can apply the formula
\begin{align}
    \delta f(x,\:y,\:...) = \sqrt{ \left(\frac{\partial f}{\partial x}\right)^2 \delta x^2 + \left(\frac{\partial f}{\partial y}\right)^2 \delta y^2 + ...}\:,
\end{align}
for the error coupling of some uncorrelated variables $x,\:y,\:...$ into a function $f(x,\:y,\:...)$ of them.
\subsection{Armlengths}\label{sec:ODErrorCouplingArmlengths}
We write the arm length between two SC as
\begin{align}
    L = \sqrt{\Delta x_{\text{along}}^2 + \Delta x_{\text{radial}}^2 + \Delta x_{\text{cross}}^2},
\end{align}
where $\Delta x_{\text{along}}$, $\Delta x_{\text{radial}}$, and $\Delta x_{\text{cross}}$ denote the position differences between these SC in the along-track, radial-track, and cross-track directions, respectively. These differences are derived from position measurements of the two SC in the corresponding direction, i.e., we get
\begin{align}
    \delta \Delta x_{\text{direction}} = \sqrt{2} \: \delta x_{\text{direction}}\label{eq:UncertaintyPositionDifference}.
\end{align}
This yields the following expression for the coupling of OD errors into arm length estimates:
\begin{widetext}
    \begin{align}
        \delta L(\Delta x_{\text{along}},\:\Delta x_{\text{radial}},\:\Delta x_{\text{cross}}) &= \sqrt{\left(\frac{\partial L}{\partial \Delta x_{\text{along}}}\:\delta \Delta x_{\text{along}}\right)^2 + \left(\frac{\partial L}{\partial \Delta x_{\text{radial}}}\:\delta \Delta x_{\text{radial}}\right)^2 + \left(\frac{\partial L}{\partial \Delta x_{\text{cross}}}\:\delta \Delta x_{\text{cross}}\right)^2},\nonumber\\
        &=\frac{\sqrt{2}}{L}\sqrt{\Delta x_{\text{along}}^2\:\delta x_{\text{along}}^2 + \Delta x_{\text{radial}}^2\:\delta x_{\text{radial}}^2 + \Delta x_{\text{cross}}^2\:\delta x_{\text{cross}}^2}.
    \end{align}
\end{widetext}
\begin{figure*}
    \begin{center}
        \includegraphics[width=1\textwidth]{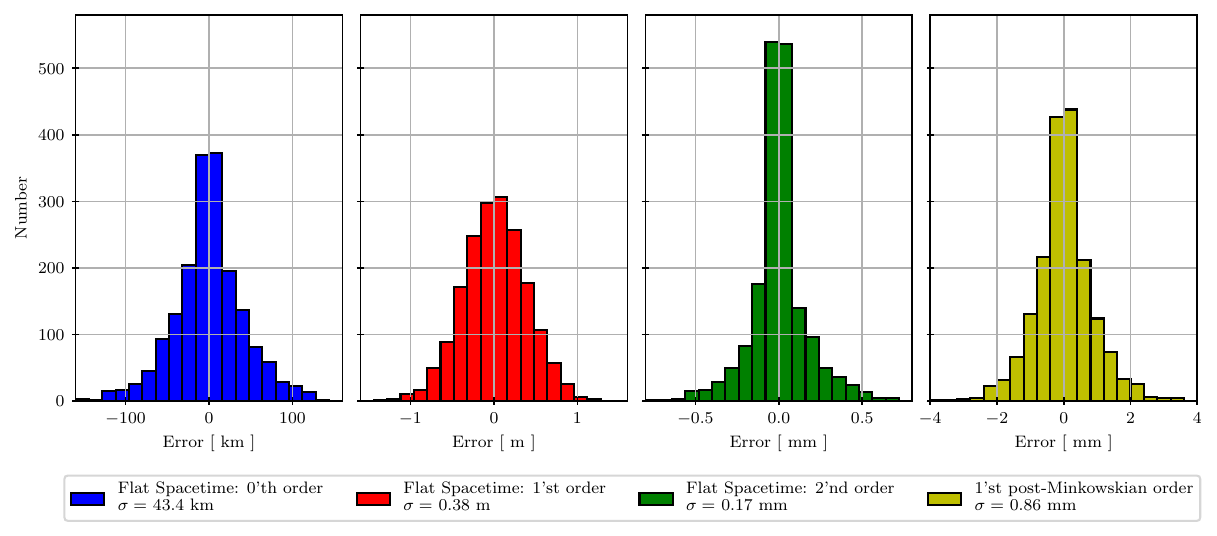}
    \end{center}
    \caption{OD error coupling into the different LTT constituents (we consider link 12). From left to right: Flat space-time order 0 (arm length), flat space-time order 1, flat space-time order 2, first post-Minkowskian order (Shapiro delay).}\label{fig:ODErrorCoupling}
\end{figure*}
To derive minimal and maximal error coupling, we evaluate this expression for arms oriented in line-of-sight and cross-track directions, respectively:
\begin{align}
    \delta L_{\text{min}} &= \delta L(L,\:0,\:0) = \sqrt{2}\:\delta x_{\text{along}}\approx \SI{3}{\kilo\m},\\
    \delta L_{\text{max}} &= \delta L(0,\:0,\:L) = \sqrt{2}\:\delta x_{\text{cross}}\approx \SI{70}{\kilo\m}.
\end{align}
Similarly, the coupling of OD errors into the first and second-order arm length time derivatives can be found to be in the order of \SI{10}{\centi\m\s\tothe{-1}} and \SI{0.1}{\micro\m\s\tothe{-2}}, respectively.
\subsection{Light Time Corrections}\label{sec:ODErrorCouplingLTC}
The LTC for the link from SC $j$ to SC $i$ is given by
\begin{align}
    \Delta_{ij} = \frac{1}{c}\: \vec{L}_{ij}\cdot \dot{\vec{x}}_{j} + O(c^{-2}),
\end{align}
where we consider the LTC in meters in contrast to \cref{eq:LightTimeCorrectionFromOrbitDetermination} in order to compare the OD error coupling between arm length and LTC estimates. To assess the OD error coupling, we rewrite this expression as
\begin{align}
    \Delta_{ij} =  \frac{1}{c}\: \left( \Delta x_{\text{along}}\: \dot{x}_{\text{along}} + \Delta x_{\text{radial}}\: \dot{x}_{\text{radial}} + \Delta x_{\text{cross}}\:\dot{x}_{\text{cross}} \right),
\end{align}
where $\dot{x}_{\text{direction}}$ denotes the absolute velocity of the sending SC in the corresponding direction. The OD error coupling can then be calculated to be
\begin{widetext}
\begin{align}
    \delta \Delta (\Delta x_{\text{along}},\:\Delta \dot{x}_{\text{along}},...)=\frac{\sqrt{2}}{c}\:\sqrt{\sum_{i \in \{ \text{along},\:\text{radial},\:\text{cross}\}} \left(\dot{x}_i \: \delta x_i\right)^2 + \left(\Delta x_i \: \delta \dot{x}_i\right)^2},
\end{align}
\end{widetext}
where we applied \cref{eq:UncertaintyPositionDifference}. Note that $\dot{x}$ in the first term refers to the absolute velocity of the sending SC. In the along-track direction, this can be approximated by the average orbital SC velocity of about \SI{30}{\kilo\m\s\tothe{-1}}. In radial-track and cross-track directions, the absolute SC velocity is significantly smaller. These terms can be neglected due to the $c^{-1}$ factor in front of the square root. The second term depends on the link orientation. To assess minimal and maximal error coupling, we consider the expression for links oriented along-track and cross-track, respectively:
\begin{align}
    \delta \Delta_{\text{min}} &= \frac{\sqrt{2}}{c}\:\sqrt{\left(\dot{x}_{\text{along}} \: \delta x_{\text{along}}\right)^2 + \left(L \: \delta \dot{x}_{\text{along}}\right)^2}\\
    &=\SI{0.3}{\m},\\
    \delta \Delta_{\text{max}} &= \frac{\sqrt{2}}{c}\:\sqrt{\left(\dot{x}_{\text{along}} \: \delta x_{\text{along}}\right)^2 + \left(L \: \delta \dot{x}_{\text{cross}}\right)^2}\\
    &=\SI{0.7}{\m}.
\end{align}
It is remarkable that the ODs, despite their high uncertainties, yield light time correction estimates at submeter accuracy.

\begin{figure*}
    \begin{center}
        \includegraphics[width=1\textwidth]{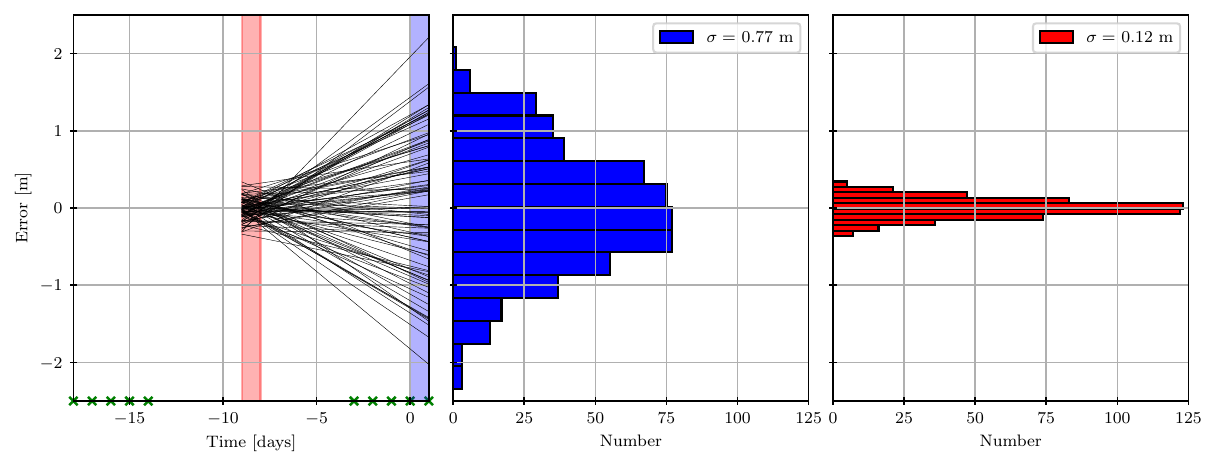}
    \end{center}
    \caption{The MOC-TC error coupling into the pseudorange evaluated for 500 realizations of 10 MOC-TCs. Left plot: Timestamps of the 10 MOC-TCs (green crosses). Error coupling time series of every fifth of the 500 realizations (black lines). Central plot: Histogram plot of the error coupling averaged over the last day (blue-shaded area in the left plot) for near-real-time applications. Right plot: Histogram plot of the error coupling averaged over day 10 (red-shaded area in the left plot) reflecting the case of available posterior data.}\label{fig:MOC-TC-error-coupling}
\end{figure*}
\section{Higher order light time corrections}\label{sec:2ndOrderLTC}
In this section, we numerically assess the coupling of OD errors into the different LTT constituents. Particularly, we here consider the LTT terms of order $O(c^{-3})$. Including these constituents, the LTTs denote \cite{Bayle:LisaOrbits,Hees:LTT-Decomposition,Chauvineau:LTT-Decomposition}
\begin{align}
    d_{ij} =&\: \frac{1}{c}\: L_{ij} + \frac{1}{c^2}\: \vec{L}_{ij}\cdot \dot{\vec{x}}_{j}\\
    +&\:\frac{L_{ij}}{2\:c^3}\left(\dot{\vec{x}}_{j}^2 + \left(\frac{\vec{L}_{ij}\cdot \dot{\vec{x}}_{j}}{L_{ij}}\right)^2 - \vec{L}_{ij}\cdot \ddot{\vec{x}}_{j}\right)\label{eq:FlatSpacetimeOrder2}\\
    +&\:\frac{2\:\mu}{c^3}\:\log \frac{\vert\vec{x}_{i}\vert + \vert\vec{x}_{j}\vert + L_{ij}}{\vert\vec{x}_{i}\vert + \vert\vec{x}_{j}\vert - L_{ij}} + O(c^{-4}).\label{eq:ShapiroDelay}
\end{align}
This expression can be understood as a post-Minkowskian LTT expansion up to the first order. The upper two lines comprise the flat space-time part (zeroth order in the post-Minkowskian expansion), which itself is an iterative solution of the implicit LTT equation: The first line includes the zeroth and first-order contribution, the second line the second-order correction. The bottom line is the first post-Minkowskian order, also known as the Shapiro delay. $\mu = \text{G}\:\text{M}_{\odot}$ denotes the gravitational parameter.\par
We numerically assess the coupling of OD errors into the different LTT constituents: Using LISA ground tracking, we simulate five years of ODs with a sampling time of one day based on an orbit file provided by ESA \cite{Bayle:LisaOrbits}. Here, we simulate uncorrelated errors for subsequent ODs to properly evaluate the OD error coupling. We use the ODs to set up the four LTT constituents. In \cref{fig:ODErrorCoupling}, we present their deviations from the actual values derived from the reference orbit file. We consider link 12, but the results hold equally for the other links as we simulate five years of orbit data, thus averaging over the link orientation. The two histograms on the left show the OD error coupling into the arm length ($\sigma = \SI{43.4}{\kilo\m}$) and the first-order flat space-time correction ($\sigma = \SI{0.38}{\m}$). The results agree with the values derived in \cref{sec:ODErrorCoupling}. The two histograms on the right show the OD error coupling into the $O(c^{-3})$ terms, i.e., the second-order flat space-time correction ($\sigma = \SI{0.17}{\milli\m}$) and the first post-Minkowskian contribution (Shapiro delay) ($\sigma = \SI{0.86}{\milli\m}$). The OD uncertainties influence these terms in the order of millimeters. Hence, the $O(c^{-3})$ terms can be accurately constructed from the ODs.

\begin{figure*}
    \begin{center}
        \includegraphics[width=1\textwidth]{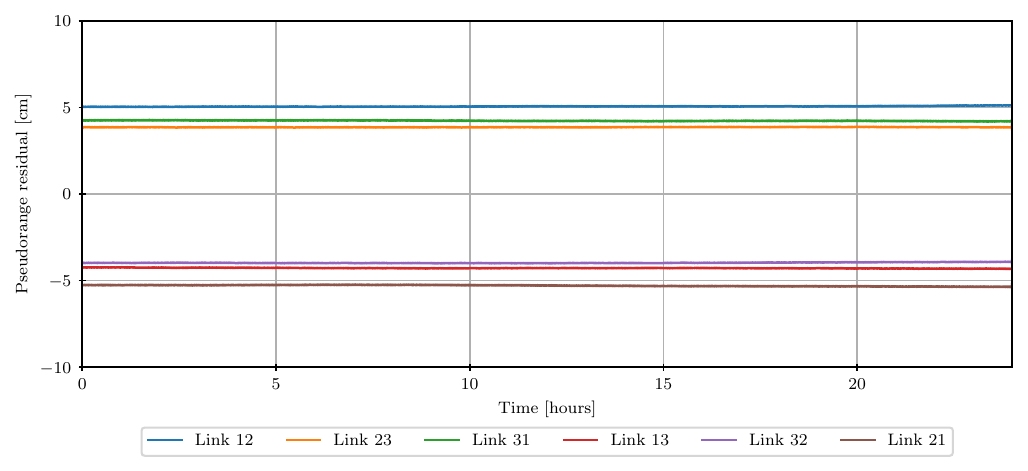}
    \end{center}
    \caption{Residual KF pseudorange estimates for the 6 links. This plot is based on the 24 hour dataset, which is used in \cref{sec:Results} to create \cref{fig:ArmlengthsEstimatesOneRealization,fig:EstimatedDifferentialScetOffsetOneRealization}.}\label{fig:Correlations}
\end{figure*}
\section{MOC-TC error coupling}\label{sec:MOC-TC-error-coupling}
The coupling of MOC-TC errors into the pseudoranges can be computed to be (see \cref{eq:MOC-TC-error-coupling})
\begin{align}\label{eq:MOC-TC-error-coupling-App}
    c \: \delta R^t_{ij}(\delta\dot{\hat{\tau}}_{1}^{t}) = c \: d_{ij}^t \cdot \delta \left(\delta\dot{\hat{\tau}}_{1}^{t}\right).
\end{align}
We assess this coupling numerically with a Monte Carlo simulation. We simulate 500 sets of 10 MOC-TCs for SC 1 according to the alternating LISA communication schedule: 5 days SC 1, 10 days the other two SC, and 5 days SC 1 (the green crosses in \cref{fig:MOC-TC-error-coupling} indicate the MOC-TC sampling times). These simulations are based on the LISA clock model (see \cref{eq:Clock-model}) and simulated SC proper time deviations from TCB (see \cref{eq:TPSwrtTCB}). We perform a second-order polynomial fit of the simulated MOC-TCs and take the time derivative. We compare the results with simulated true MOC-TCs to which we apply the same procedure.\footnote{This analysis does not reflect fit imperfections, since we apply the same fit to the MOC-TCs and to the true reference MOC-TCs.} Taking the difference yields estimates for $\delta \left(\delta\dot{\hat{\tau}}_{1}^{t}\right)$. Scaling this with the average LTT yields the MOC-TC error coupling into the pseudorange.\par
The simulation results can be seen in \cref{fig:MOC-TC-error-coupling}: The plot on the left shows the error coupling time series of the 500 realizations (see black lines). For near-real-time applications (see blue-shaded area), the error coupling is almost one order of magnitude higher than in the case of available posterior data (see red-shaded area). We do not plot the time series further to the left, where the error coupling increases again, since previous MOC-TCs are not taken into account (we just consider the 10 MOC-TCs indicated by the green crosses in \cref{fig:MOC-TC-error-coupling}); therefore, the MOC-TC error coupling in this region does not reflect reality, where previous MOC-TCs would be available. The histogram plot in the center shows the error coupling averaged over the last day of the data set corresponding to the case of near-real-time pipelines. Here MOC-TC errors couple at submeter order ($\sigma = \SI{0.77}{\m}$), which is sufficient considering the accuracy requirement for the pseudorange disentanglement. The histogram plot on the right shows the error coupling averaged over day 9. This corresponds to the case of available posterior data, which increases the accuracy of the second-order polynomial fit so that the error coupling is reduced by almost one order of magnitude ($\sigma = \SI{0.12}{\m}$).

\section{Correlations between LTTs and SCET desynchronizations}\label{sec:Correlations}
Errors in the ODs and the MOC-TCs limit the estimation accuracy for the LTTs ($\sigma\approx\SI{80}{\centi\m}$) and the differential SCET desynchronizations ($\sigma\approx\SI{30}{\centi\m}$) (see \cref{sec:Results}). However, the errors in these estimates are correlated. To illustrate this we apply the observation model (see \cref{eq:ObservationModel}) to the KF state vector estimates after two iterations. Thus, we obtain KF pseudorange estimates
\begin{align}
    \hat{R}_k^{+} = h(\hat{x}_k^{+},\:p_k,\:0).
\end{align}
We compare these to the true pseudoranges (obtained as validation support from LISA Instrument), which we transform to TCB. The residuals are plotted in \cref{fig:Correlations}. It can be seen that the combined error of LTTs and differential SCET desynchronizations is in the order of \SI{5}{\centi \m} one order of magnitude below the individual errors. When using the estimates for the LTTs and the differential SCET desynchronizations in TDI, we are limited by the combined errors and not by the individual ones.

\clearpage
\bibliography{references}
\end{document}